\documentclass[twocolumn,preprintnumbers,amsmath,amssymb,aps,prb,longbibliography]{revtex4-1}
\usepackage{graphicx}
\begin{document}

\title{Dynamics of Magnus Dominated Particle Clusters, Collisions, Pinning and Ratchets}
 
\author{C. Reichhardt and C. J. O. Reichhardt}
\affiliation{Theoretical Division and Center for Nonlinear Studies,
Los Alamos National Laboratory, Los Alamos, New Mexico 87545, USA}

\date{\today}

\begin{abstract}
Motivated by the recent work in skyrmions and active chiral matter systems, 
we examine pairs and small clusters of repulsively interacting point particles in 
the limit where the dynamics is dominated by the Magnus force. 
We find that particles with the same Magnus force can form stable pairs, triples 
and higher ordered clusters or exhibit chaotic motion. 
For mixtures of particles with opposite Magnus force,
particle pairs can combine to form translating dipoles. 
Under an applied drive,
particles with the same Magnus force translate;
however, particles with different or opposite
Magnus force exhibit a drive-dependent decoupling transition. 
When the particles interact with a repulsive obstacle, they can form localized
orbits with depinning or unwinding transitions under an applied drive.
We examine the interaction of these particles
with clusters or lines of obstacles, and find that the particles can become trapped in 
orbits that encircle multiple obstacles. 
Under an ac drive, we observe a series of ratchet effects,
including ratchet reversals, for particles interacting
with a line of obstacles due to the formation of commensurate orbits.
Finally, in assemblies of particles with mixed Magnus forces of the same sign,
we find that the particles with the largest Magnus force become localized
in the center of the
cluster, while for mixtures with opposite Magnus forces, the motion is
dominated by transient 
local pairs or clusters,
where the translating pairs can be regarded as a form of active matter.
\end{abstract}

\maketitle

\vskip 2pc

\section{Introduction}
There are a variety of systems that can be described
as local clusters of interacting particles, including
colloids \cite{Bubeck99,Drocco03,Perry15},
Coulomb clusters \cite{Bedanov94,Nelissen06},
vortices in type II superconductors \cite{Cabral04,Komendova13},
dusty plasmas \cite{Juan98a},
Wigner crystals \cite{Bhattacharya16}, vortices
in superfluids \cite{Seo17,Gauthier19},
skyrmions \cite{Zhao16,Schaffer19},
granular matter \cite{Lim19},
and active matter assemblies \cite{Schmidt19}.
In many of these systems, the cluster formation arises
when the particles experience a local confinement
or self-trapping due to the nature of the pairwise particle-particle
interactions \cite{CostaCampos13,Reichhardt07}.
Under various types of driving, these systems can 
exhibit interesting dynamical effects
including self assembly \cite{Niu17,Abdi18},
rotating gear behavior \cite{Aubret18,Williams16},
and depinning phenomena \cite{Reichhardt17}.
In most of these systems,
the dynamics is overdamped;
however, some systems also include
nondissipative effects such as inertia or
Magnus forces.
In particular, Magnus forces produce a velocity component
that is perpendicular to the net force experienced by a particle,
and such forces arise for
vortices in fluids \cite{Aref88,Boatto99,Aref07,Reinaud18},
active spinners \cite{Grzybowski01,Climent07,Banerjee17,Denisov18,Gorce19},
chiral active matter \cite{Chepizhko19},
charged particles in magnetic fields \cite{Schirmacher15},
and skyrmions in chiral magnets \cite{Muhlbauer09,Yu10,Nagaosa13}.
One consequence of this
is that pairs or clusters of particles can
undergo rotations or spiraling motion when they
enter a confining potential 
\cite{Liu13,Muller15,Buttner15,Reichhardt15a,Reichhardt15}
or are subjected to a quench \cite{Brown18}. 
If damping is present, these spiraling motions are transient unless there is some form
of external driving.
Less is known about 
how Magnus-dominated particles
would interact with obstacles or pinning sites; however, 
there are some  studies which indicate that the Magnus force strongly modifies the 
dynamics compared to overdamped systems
\cite{Reichhardt15a,Reichhardt15,Brown18,Jiang17,Legrand17,Kim17,Litzius17,Juge19,Zeissler19,Feilhauer19}. 

Motived by our previous work on point particle models of skyrmions
interacting with each other and
with random \cite{Reichhardt15,Lin13} or periodic pinning  \cite{Brown18}, 
where the particles have both a Magnus and a damping force,
we consider the limits of zero 
damping or very low damping
and study the Magnus-dominated dynamics of pairs and
small clusters of particles interacting with each other
and with pinning sites.
We consider mixtures with identical Magnus forces,
dispersion in the Magnus force,
and assemblies with opposite Magnus forces. 
In the case of a pair of particles
with the same sign and magnitude of the Magnus force,
we find that a bound rotating pair 
forms despite the repulsive particle-particle interactions,
and that under an external drive the pair remains coupled and translates 
at $90^\circ$ with respect to the driving direction.
If the magnitude of the Magnus force is not the same for both particles,
a drive dependent decoupling transition occurs.
For higher numbers of particles we find various types of stable rotating
states, including rotating pairs that rotate around each either. 
For larger clusters we observe chaotic dynamics in which
the system breaks up into smaller clusters with some particles
jumping from one cluster to another.
When the Magnus forces
of a pair of particles are of the same magnitude but different sign,
the particles
form a dipole which translates in a direction determined by
the initial orientation of the pair, with dipoles of smaller size translating more rapidly.
If the Magnus forces are of different sign and magnitude,
the particles form a translating pair that can break apart and reform if a collision
with an obstacle or other particles occurs.
We argue that assemblies of particles with mixed Magnus force sign represent
a new example of an active matter system.  

When repulsive obstacles are added to the system,
we find that the particles can form stable bound circulating orbits around the obstacles
and exhibit a depinning transition under an applied drive.
We show that it is possible for pairs and clusters of
particles to collide with and become localized by a obstacle.   
If damping is present, these pinned states are transient and the particle
gradually winds away from the obstacle.
A particle interacting
with a cluster of defects
can enter an orbit that encircles all of the defects.
In the overdamped limit, an asymmetric cluster of defects produces a
diode-like effect for driving in different directions, but in the Magnus-dominated
limit this diode effect disappears and the particles circle around the entire
cluster.
A particle driven toward a line of obstacles experiences a
Magnus force-induced deviation in its
direction of motion as it approaches the line until it breaks through the line,
and this deviation is reduced for increased driving force.
We also find that it is possible to produce a ratchet
effect for a particle that is placed by a line of obstacles
when a biharmonic ac drive is applied.  Here the particle can form circular orbits that
create a gear-like motion when combined with the periodicity of the line of obstacles.
Reversals in the ratchet current occur as a function of ac amplitude and Magnus force.  
Finally, we examine the chaotic dynamics of smaller clusters
and show that if there is dispersion in the Magnus force, the particles
with the largest Magnus forces become localized in the center of the cluster. 
 
Our results should be relevant for
skyrmions in the absence of damping or
in the low damping limit in the presence of a drive,
for certain models of point vortex dynamics in superfluids or Bose-Einstein condensates with
fluid flows, and for active spinners and active chiral colloidal systems.  

\section{Simulation}
We consider a two-dimensional system
with periodic boundary conditions in the $x$ and $y$-directions
containing
$N$ particles that are initially
placed at fixed distances from each other.
Typically we use initial conditions in which the particles are in one-dimensional lines.  
The dynamics of particle $i$ are governed by the following
undamped equation of motion:
\begin{equation} 
\alpha^i_m {\hat z} \times {\bf v}_{i} =
{\bf F}^{pp}_{i} + {\bf F}^{obs} +  {\bf F}^{D},
\end{equation}
where ${\bf v}_i$ is the velocity of particle $i$ and
$\alpha^i_{m}$ is the coefficient of the Magnus term,
which creates a velocity component perpendicular to the net applied forces.
Each particle can be assigned a different amplitude or sign of
$\alpha^i_m$.
The particle-particle interaction force is given by
${\bf F}^{pp}_{i} = \sum^{N}_{j=1}K_{1}(r_{ij}){\hat {\bf r}_{ij}}$,
where $r_{ij} = |{\bf r}_{i} - {\bf r}_{j}|$ is the distance
between particles $i$ and $j$, $\hat {\bf r}_{ij}=({\bf r}_i-{\bf r}_j)/r_{ij}$, and
the modified Bessel function $K_{1}(r)$ falls
off exponentially for large $r$.
This form of the interaction was previously used in particle-based
models of skyrmions in two-dimensional systems
\cite{Reichhardt15a,Reichhardt15,Brown18,Lin13}.  
The driving force ${\bf F}^{D}=F^D{\bf {\hat x}}$ is applied uniformly to all
particles.
An individual particle in the Magnus force-dominated limit
moves at $90^\circ$ with respect to the driving force,
so that when the drive is applied in the $x$ direction,
the particle moves in the $y$ direction. 
The term ${\bf F}^{obs}=\sum_{k=1}^{N_p}$ represents the force from
$N_p$ obstacles, which take the form of particles that are permanently fixed
in place.
In some cases,
we add a damping term $\alpha^i_{d}{\bf v}_{i}$
to the equation of motion which aligns the velocities
in the direction of the external forces.
Under a drive, a particle experiencing both Magnus and damping
forces
moves at an angle $\theta = \arctan(\alpha_{m}/\alpha_{d})$. 
We measure the particle velocities both parallel,
$\langle V_{x}\rangle=N^{-1}\sum_{i}^{N}{\bf v}_i \cdot {\bf \hat x}$,
and perpendicular,
$\langle V_{y}\rangle=N^{-1}\sum_{i}^{N}{\bf v}_i \cdot {\bf \hat y}$
to the drive.

\section{Dynamics of Coupled Particles}

\begin{figure}
\includegraphics[width=3.5in]{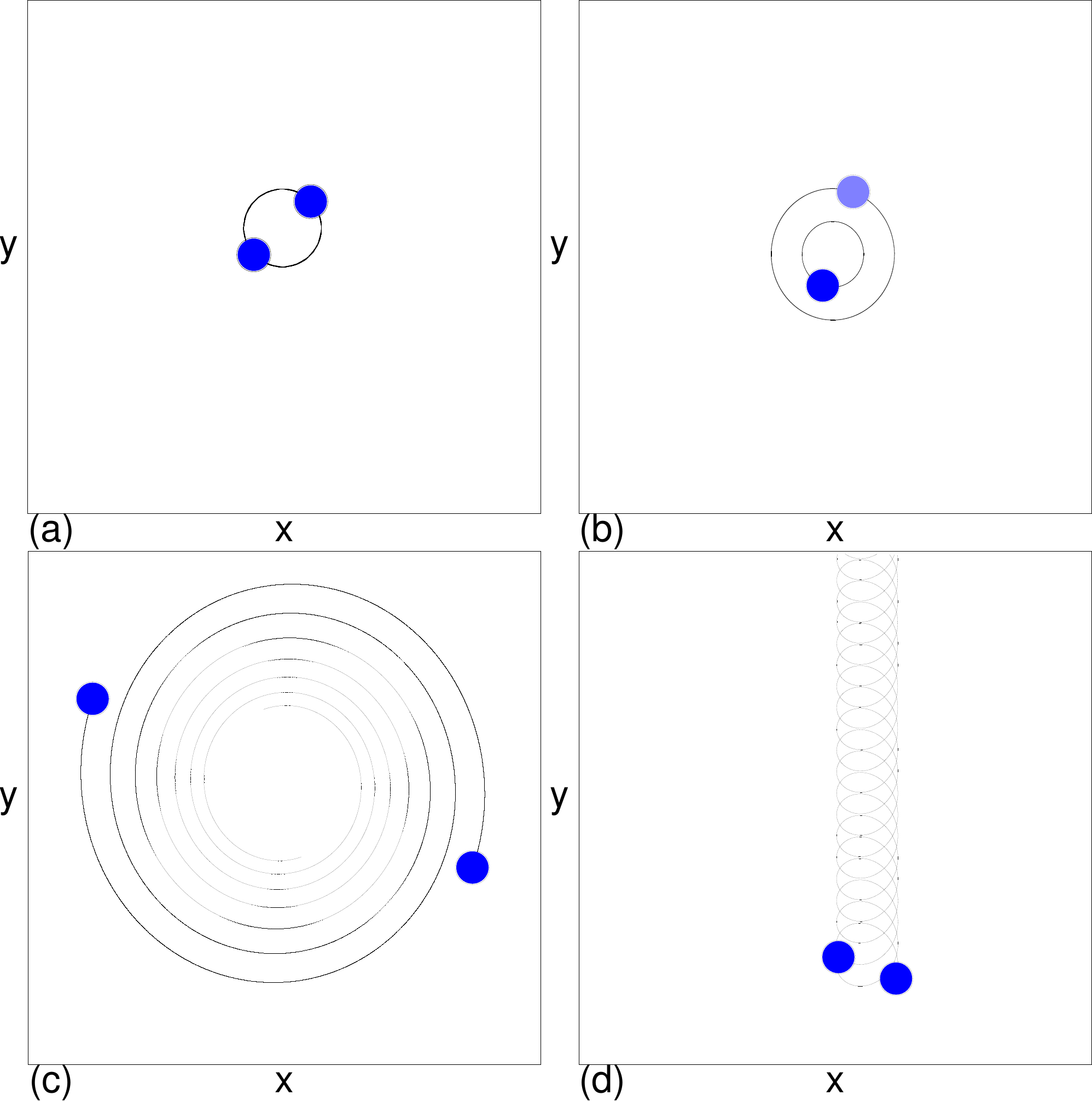}
\caption{
The particle locations (dots) and trajectories (lines)
for pairs of interacting particles. (a) When
$\alpha^{1}_{m}=\alpha^2_m = 1.0$,
the particles form a clockwise rotating bound pair. 
(b) For $\alpha^{1}_{m} = 2.0$ and $\alpha^{2}_{m} = 1.0$, the particles form
nested orbits where the particle with the higher Magnus force is closer
to the center.
(c) A system with $\alpha^1_m=\alpha^2_m=1.0$ in which a finite
damping term $\alpha_d=0.1$ has also been added, causing
the particles to spiral out gradually.
(d) The $\alpha^1_m=\alpha^2_m=1.0$ system from (a)
with an additional drift force $F_{D} = 0.075$ applied in the $x$-direction, causing  
the pair to translate in the negative $y$-direction. }
\label{fig:1}
\end{figure}

We first consider particles with the same sign and
magnitude of the Magnus force. 
In Fig.~\ref{fig:1}(a) we show an image of
the trajectories of two particles with
$\alpha^1_m=\alpha^2_m=1.0$ initialized a distance $R$ apart.
In an overdamped system, the particles would move away from each other,
but here they form 
a pair and rotate around each other in a clockwise manner.
The particles remain confined to the pair due to
the Magnus force which
generates velocities that are perpendicular to the net forces on each particle.
When $\alpha^1_m \neq \alpha^2_m$,
the particles form a nested pair as illustrated in Fig.~\ref{fig:1}(b) for
$\alpha^{1}_{m} = 1.0$ and $\alpha^{2}_{m} = 2.0$,
with the larger Magnus force particle orbiting closer to the center.
If we add a finite damping term  of $\alpha_{d} = 0.1$
to the $\alpha^1_m=\alpha^2_m=1.0$ system in
Fig.~\ref{fig:1}(a), the particles gradually spiral away from each other
as shown in Fig.~\ref{fig:1}(c),
and in the long time limit, the presence of damping eventually causes the particles
to come to a standstill.
If only one particle has damping, the overall motion still damps away
since the damped particle couples to the undamped particle and dissipates
its energy,
so as long as there is some damping in the system both
particles will eventually come to rest unless an external drive is applied.
In the zero damping limit, when there is an applied drive
the rotating pair remains coupled and its center of mass
translates, as shown in Fig.~\ref{fig:1}(c) for the $\alpha^1_m=\alpha^2_m=1.0$ system
under a drive of $F_{D} = 0.075$.
The $x$ direction drive causes
the pair to translate in the negative $y$-direction,
giving a skyrmion Hall angle of $90^\circ$.
Here the intrinsic skyrmion Hall angle
is defined as $\theta_{sk}^{\rm int} = \arctan(\alpha_{m}/\alpha_{d})$
\cite{Reichhardt15,Jiang17,Litzius17}.  
In the presence of damping, the driven pair in
Fig.~\ref{fig:1}(d) gradually spiral away from
each other and translate separately
at a Hall angle less than $90^\circ$. 

\subsection{Systems with opposite Magnus force}

\begin{figure}
\includegraphics[width=3.5in]{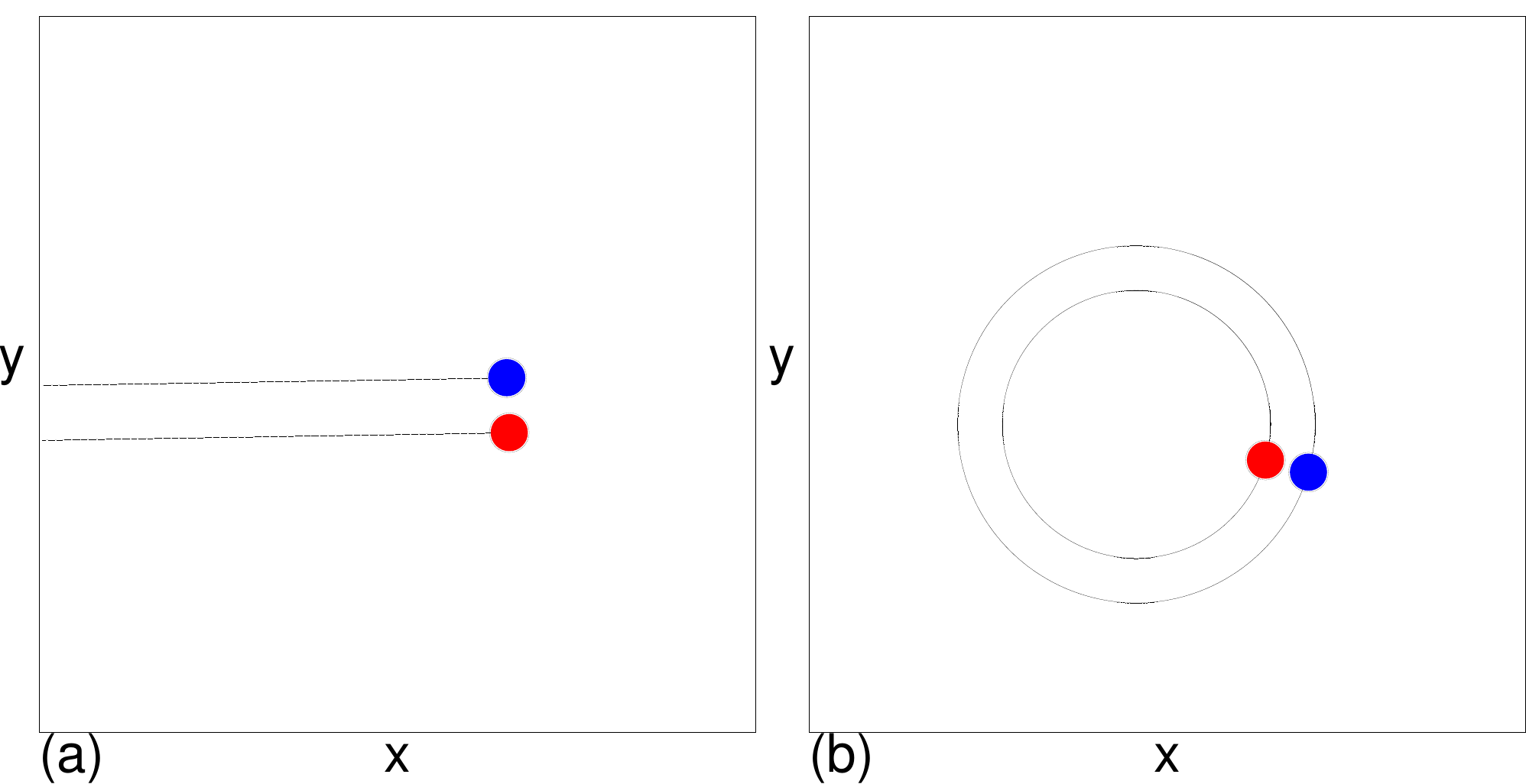}
\caption{
  The particle locations (dots) and trajectories (lines) for pairs
  of interacting particles.
  (a) When $\alpha^{1}_{m} = 2.0$ and $\alpha^{2}_{m} = -2.0$,
  the particles form a dipole that translates in a fixed direction.
  (b) When $\alpha^{1}_{m} = 1.65$ and $\alpha^{2}_{m} = -2.0$,
  the dipole moves in a circular orbit. 
}
\label{fig:2}
\end{figure}

\begin{figure}
\includegraphics[width=3.5in]{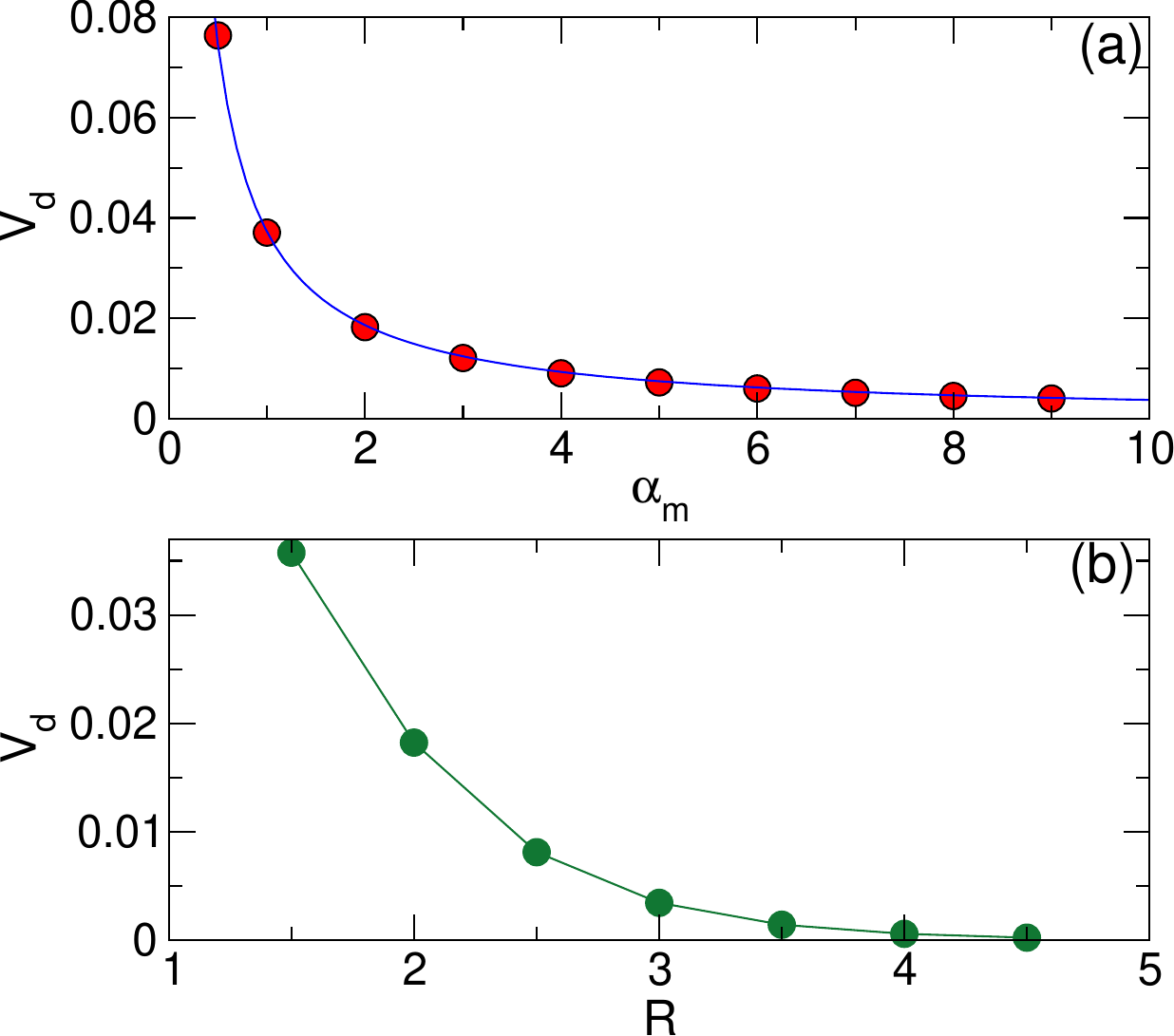}
\caption{
  The dipole velocity $V_{d}$ vs $\alpha_{m}$ for the system in
  Fig.~\ref{fig:2}(a) with $\alpha^1_m=\alpha_m$ and $\alpha^2_m=-\alpha_m$,
  where the initial distance between the particles is $R=2.0$.
  The solid blue line is a fit to $1/\alpha_{m}$.
  (b)  $V_{d}$ vs $R$ for the same system with fixed $\alpha_{m} = 2.0$.  
}
\label{fig:3}
\end{figure}

When two particles that have Magnus forces which are equal in magnitude
but opposite in sign
are brought together, they form a bound pair that translates in a fixed direction
even in the absence of an applied drive.
The repulsive interaction between the two particles produces an outwardly directed
force on each particle, and the Magnus term rotates this force by 90$^\circ$ for one
particle and by $-90^\circ$ for the other, producing a net translation instead of a
rotation.
In Fig.~\ref{fig:2}(a), a pair of particles with $\alpha^1_m=2.0$ and $\alpha^2_m=-2.0$
maintain a fixed distance from each other and translate
in a direction that is determined by the initial placement of the particles.
The speed of the dipole pair
increases as the initial distance $R$ between the particles decreases,
since the pairwise interaction force increases at smaller distances,
while the dipole drift velocity  $V_d$ 
is given by
$V_{d} \propto K_{1}(R)/\alpha_m$,
where $\alpha_m=|\alpha^1_m|=|\alpha^2_m|$.
In Fig.~\ref{fig:3}(a) we plot the
measured velocity $V_d$ versus $\alpha_m$
for the system in Fig.~\ref{fig:2}(a) at a fixed 
initial separation distance of $R=2.0$.
The solid line is a fit to $1/\alpha_{m}$.
In Fig.~\ref{fig:3}(b) 
we show $V_{d}$ versus $R$ for fixed $\alpha_{m} = 1.0$
in the same system.
The dipole velocity decreases approximately exponentially
with increasing distance
at large $R$, as expected for the function $K_{1}(R)$.     
In Fig.~\ref{fig:2}(b) we illustrate the dipole trajectory for
a system with Magnus forces of opposite sign but unequal magnitude,
$\alpha^{1}_{m} = 1.65$ and $\alpha^{2} = -2.0$, 
where the dipole curves into a localized circular orbit.
As the difference in magnitude of the Magnus forces increases,
the circular orbit becomes tighter.

\section{Dynamics Under a Drive}

\begin{figure}
\includegraphics[width=3.5in]{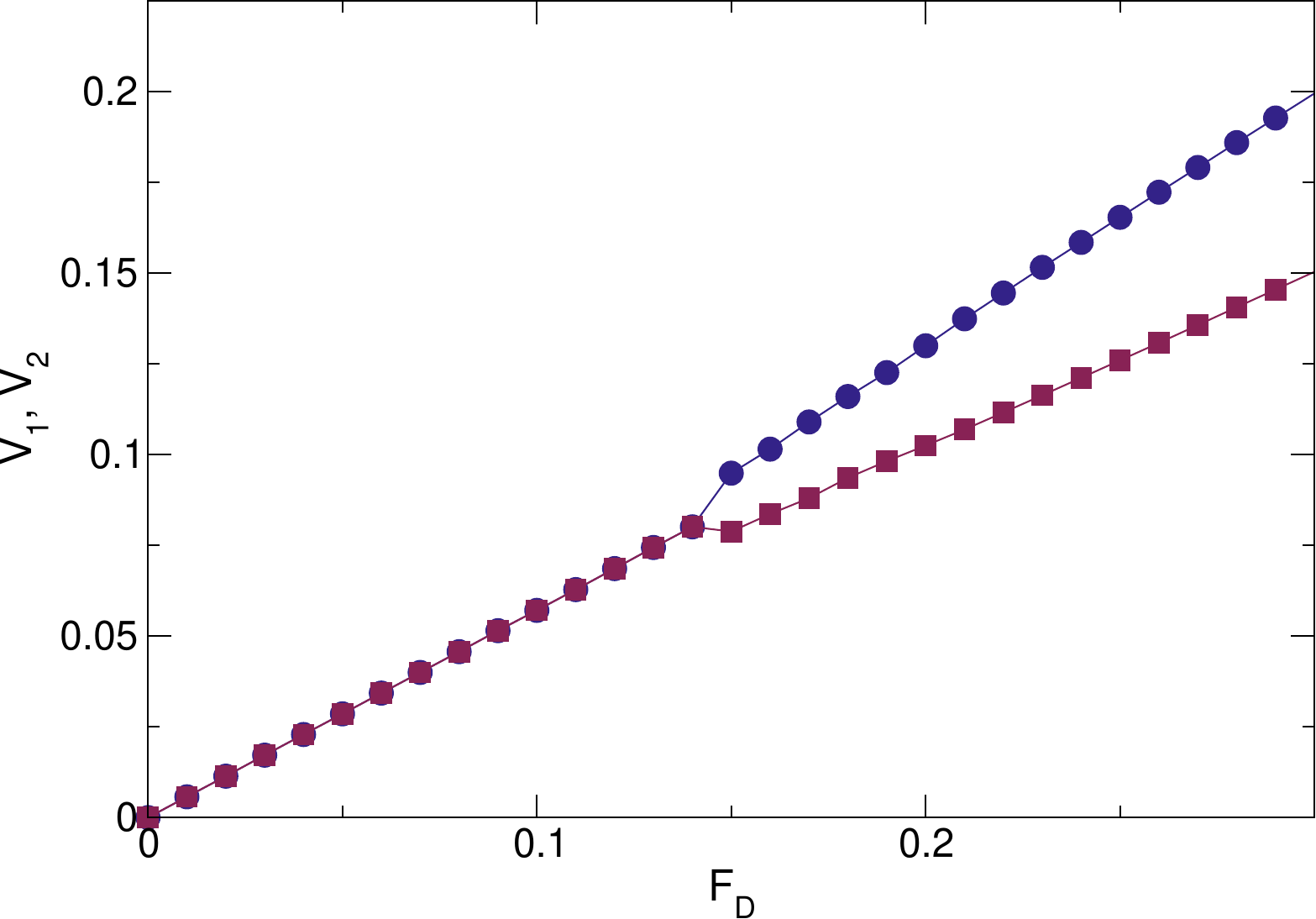}
\caption{
  The velocities $V_1$ (blue) and $V_2$ (red) of a pair of particles
  vs $F_{D}$ for a system with $\alpha^{1}_{m} = 1.6$ and
  $\alpha^{2}_{m} = 2.0$, showing a  drive induced
  decoupling transition near $F_{D} = 0.15$.  
}
\label{fig:4}
\end{figure}

We next consider the effect of applying a
driving force in the positive $x$-direction, which causes isolated particles with a
positive Magnus force to move in the negative $y$ direction.
For a pair of particles with Magnus forces of equal sign and magnitude,
the pair remains coupled when the drive is applied
and translates perpendicular to the drive, as shown in Fig.~\ref{fig:1}(d).
If the magnitude of the Magnus forces are unequal,
there is a critical driving force above which the pair decouples.
In Fig.~\ref{fig:4} we plot the
velocities $V_1$ and $V_2$ of a pair of particles versus driving force $F_{D}$
for a system with $\alpha^{1}_{m} = 1.6$ and
$\alpha^{2}_{m} = 2.0$.
For $F_D\leq 0.15$, $V_1=V_2$ and the particles are coupled into a dipole,
while for $F_{D} > 0.15$, the pair decouples as indicated by the
change in the velocities.
The critical driving force $F_c$ at which the decoupling occurs decreases
as the difference $|\alpha^1_m-\alpha^2_m|$ increases, while $F_c$ increases
as the separation $R$ decreases.

A cluster containing more than two particles that all have
the same $\alpha_{m}$ remains coupled under an applied drive,
but when some of the particles have different values of
$\alpha_{m}$, multiple decoupling transitions can occur.  

\subsection{Dynamics with Obstacles and Depinning }

\begin{figure}
\includegraphics[width=3.5in]{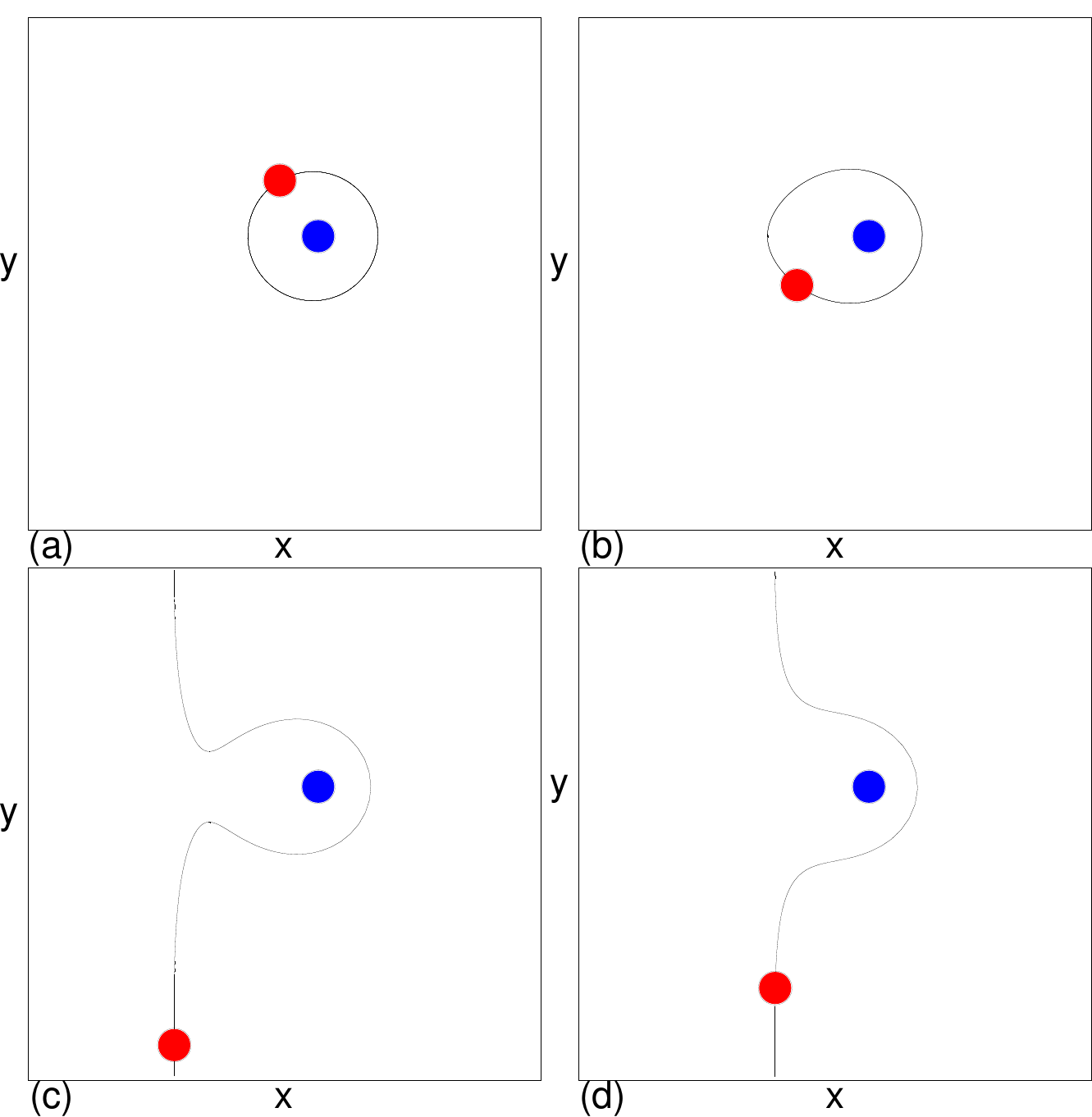}
\caption{
  The particle position (red dot) and trajectory (line) with the obstacle
  location (blue dot)
  for a single particle
  interacting with a stationary obstacle in the form of a permanently fixed
  particle.
  The particle has $\alpha_{m} = 2.0$ and is initialized at a distance $R=1.5$ from
  the obstacle, and the applied drive is 
(a) $F_{D} = 0.005$, (b) $F_{D} = 0.015$, (c) $F_{D} = 0.0165$ and (d) $F_{D} = 0.025$.  
}
\label{fig:5}
\end{figure}

\begin{figure}
\includegraphics[width=3.5in]{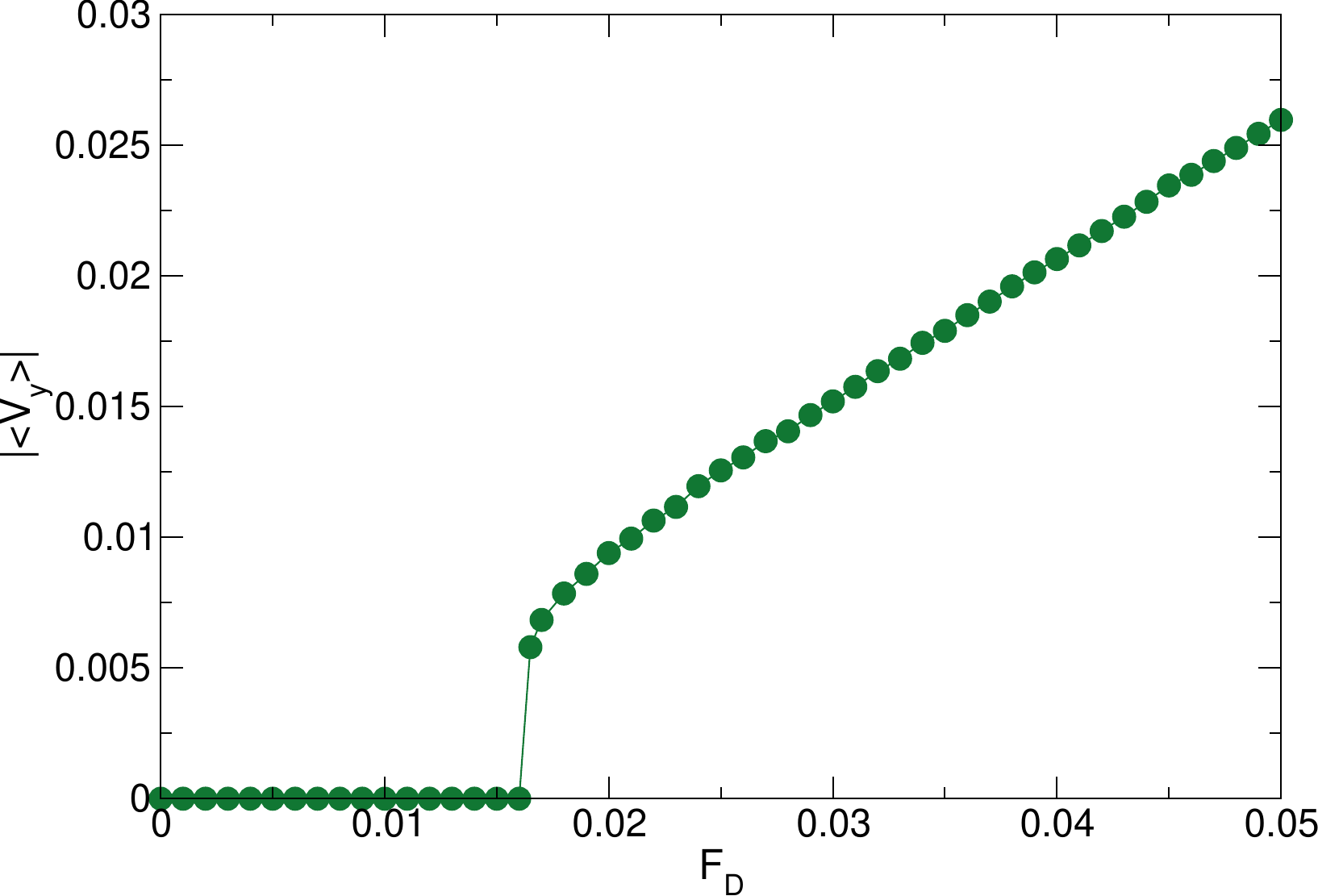}
\caption{
  $|\langle V_y\rangle|$,
  the absolute value of the average velocity in the $y$-direction
  of the particle from the system in Fig.~\ref{fig:5}, vs $F_{D}$,
showing a depinning transition at $F_{D} = 0.016$.
}
\label{fig:6}
\end{figure}

We next study the effects of driven particles interacting with a repulsive obstacle.
To begin, we consider a single particle under an applied drive
interacting with an obstacle which is modeled
as another particle that is fixed permanently in place, giving
a repulsive force  between the particle and the obstacle.
In the overdamped limit, there
is no pinning effect and the particle simply moves away from the obstacle
due to the pairwise repulsion.  
In Fig.~\ref{fig:5}(a), a particle with $\alpha_{m} = 2.0$ under a driving force
of $F_{D} = 0.005$ initialized at a distance of $R=1.5$ from the obstacle
forms a localized pinned orbit around the 
obstacle.
At $F_D=0.015$ in the same system, Fig.~\ref{fig:5}(b) indicates that
the particle is still localized but the orbit becomes distorted by the drive. 
In Fig.~\ref{fig:5}(c) at $F_{D} = 0.165$, the particle 
has depinned and translates in the $y$ direction, interacting with the obstacle during
each pass through the periodic boundary conditions.
At $F_D=0.025$ in Fig.~\ref{fig:5}(d),
the interaction with the obstacle is diminished and the pinch point
in the trajectory near the
obstacle has disappeared.
In Fig.~\ref{fig:6}(a) we plot
the absolute value of the average particle velocity in the
$y$-direction, $|\langle V_{y}\rangle|$, versus $F_{D}$.
We find a clear
region where the particle is pinned, as indicated
by $|\langle V_{y}\rangle| = 0$, along
with a critical depinning force at $F_{c} = 0.016$.
In most systems where depinning occurs,
there must be an attractive interaction between the particle and a defect so that
the particle can settle into a potential energy minimum and stop moving.
It is possible in some overdamped
systems for the particle to become trapped behind a repulsive barrier,
but even
in that case the particle
comes to 
rest and can be described as jammed \cite{Reichhardt17}.
Here we find a depinning transition in which the particle is always moving but remains
localized below depinning.
If the sign of the Magnus force is reversed, the same dynamics occurs 
but the particle depins in the opposite direction.
The depinning threshold depends on the magnitude of $\alpha_{m}$ and the 
initial distance $R$ at which the particle is placed from the obstacle.

\begin{figure}
\includegraphics[width=3.5in]{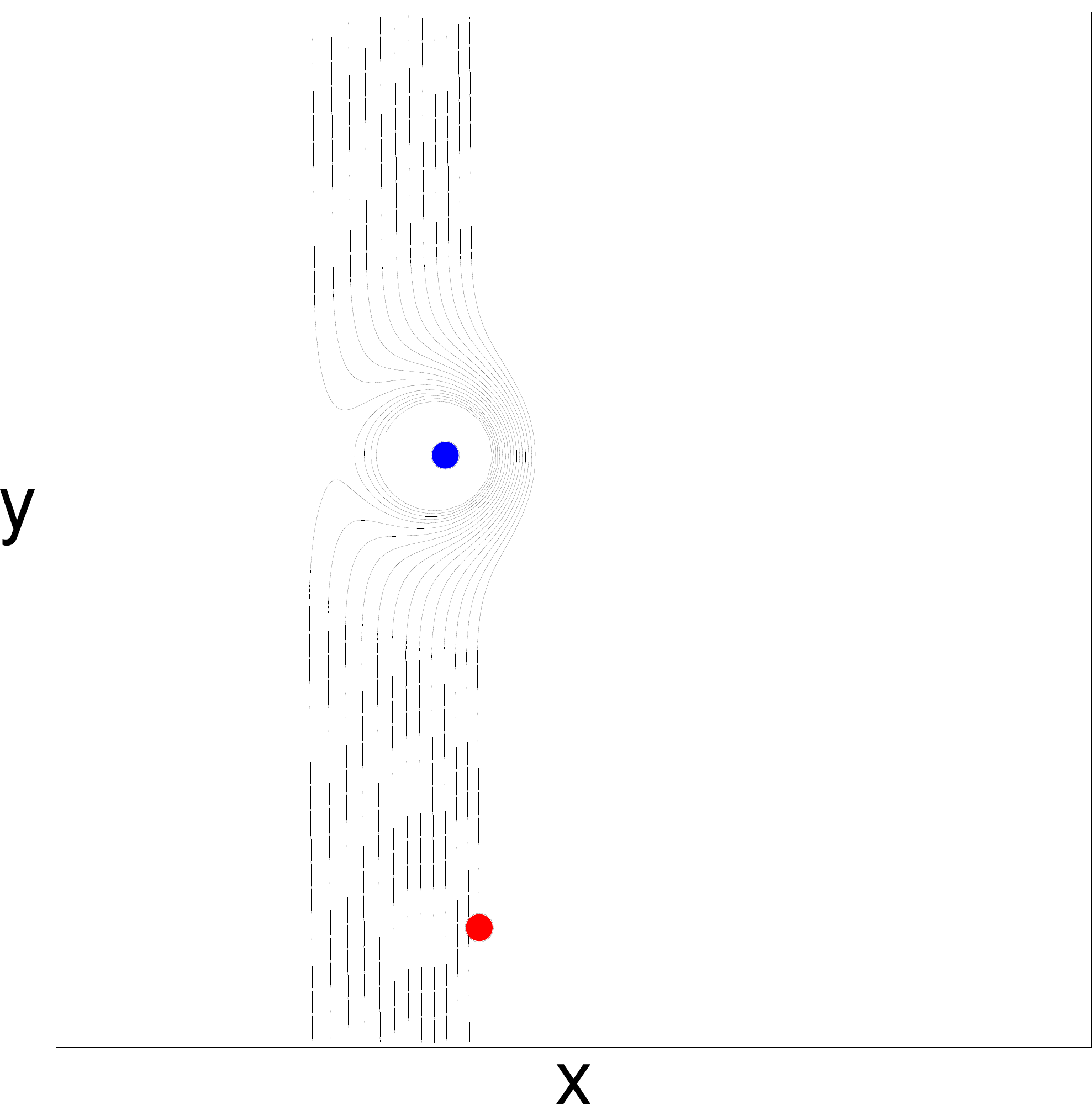}
\caption{
  The particle position (red dot) and trajectory (lines) along with the
  obstacle location (blue dot) for
  the system from Fig.~\ref{fig:5} with $\alpha_m=2.0$ and $R=1.5$ at $F_{D} = 0.01$
  where an additional damping term of $\alpha_{d} = 0.01$ has been added to
  the dynamics.  The particle gradually
spirals away from the obstacle.
}
\label{fig:7}
\end{figure}

If we add some damping to the particle dynamics,
the particle does not remain localized
but always escapes via an unwinding transition.
This process is illustrated in Fig.~\ref{fig:7}(a) 
for the system from Fig.~\ref{fig:5} with an added damping
of $\alpha_{d} = 0.01$ at a fixed drive of $F_{D} = 0.01$, which is
below the threshold depinning force found in
Fig.~\ref{fig:6} for the undamped system.
When damping is present,
after each orbit
the particle gradually 
moves away from the obstacle until eventually it depins
and then translates at an angle $\theta_{sk} = \arctan(\alpha_{m}/\alpha_{d})$.  
If the interaction between the particle and the obstacle
is attractive, when damping is present
the particle gradually spirals into the obstacle,
while if a drive is also applied,
the particle spirals inward until it reaches an equilibrium point
at which the driving force balances the attractive force from the obstacle.

\begin{figure}
\includegraphics[width=3.5in]{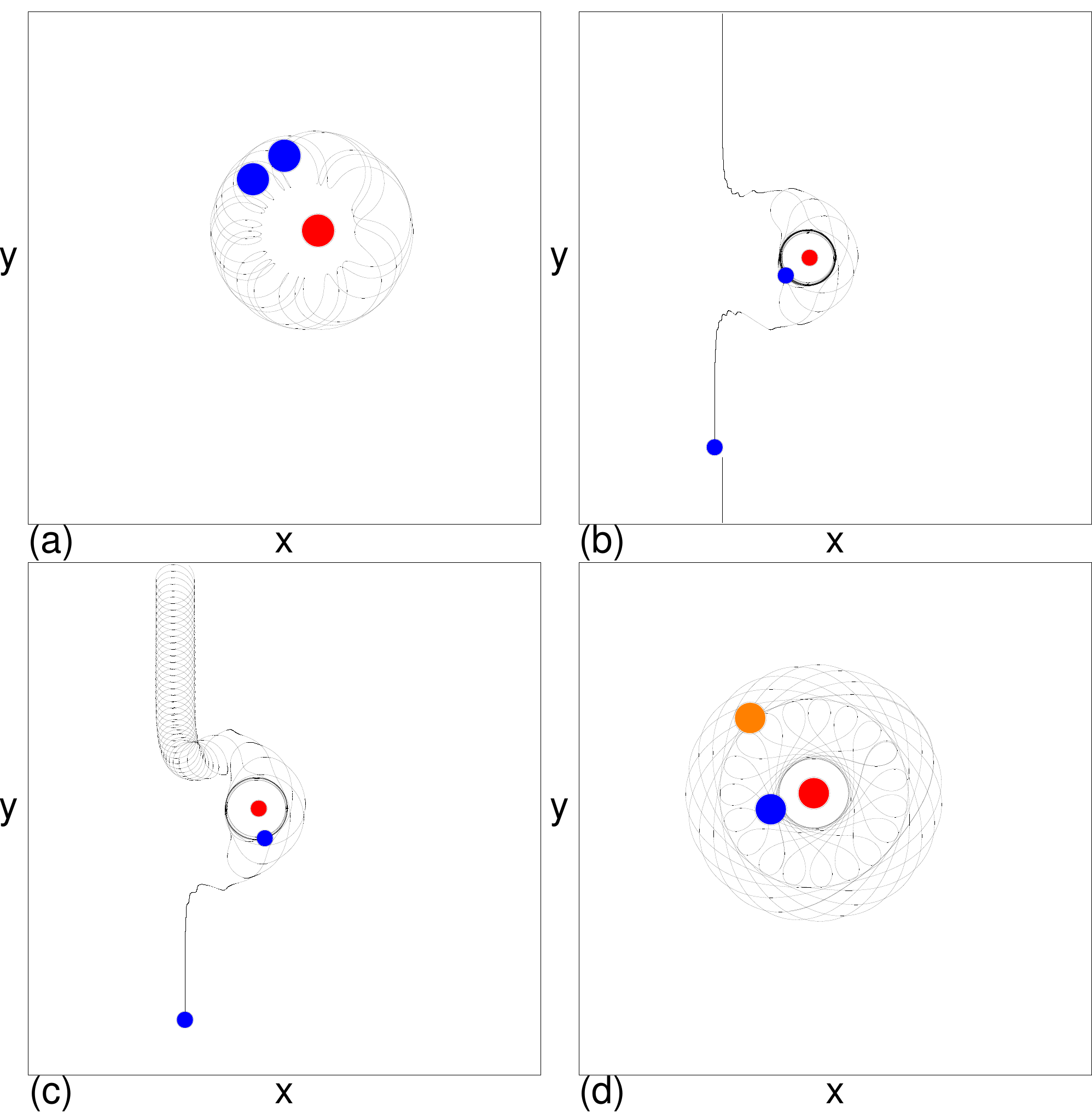}
\caption{
  The particle positions (blue and orange dots) and trajectories (lines) along with the
  obstacle position (red dot).
  (a) Two particles trapped at an obstacle for
  $\alpha_{m}^1=\alpha_m^2 = 2.0$ at $F_{D} = 0.005$ and $R=1.5$.
  (b) The same system at $F_{D} = 0.01$ where only one particle can be trapped.
  (c) The same system at $F_D=0.01$ in which the two particles are initially
  in a rotating pair that collides with the defect, which traps one of the particles.
  (c) Two particles trapped at an obstacle for
  $\alpha_m^1=2.0$ and $\alpha_m^2=-2.0$ at $F_D=0.005$ and $R=1.5$,
  where the Magnus forces of the particles have opposite signs.
}
\label{fig:8}
\end{figure}

A single repulsive obstacle can also capture multiple particles.
An example of this process appears
in Fig.~\ref{fig:8}(a) for a sample with two particles where
$\alpha_m^1=\alpha_m^2=2.0$, $R=1.5$, and $F_D=0.005$,
where the two particles form a pair that rotates around the obstacle.
Due to the applied drive, the trajectories are denser
on the left side
of the obstacle.
When the drive is increased, a depinning transition
occurs in which one particle depins while
the other remains localized,
as shown in Fig.~\ref{fig:8}(b) for the same system at $F_{D} = 0.01$. 
Due to the periodic boundary conditions, the depinned particle returns
and interacts with the obstacle again, passing through
a spiraling orbit before escaping.
At a higher drive of $F_{D} > 0.015$, the second particle
also depins.  
If the two particles are initially in a pair away from the obstacle,
then when they collide with the obstacle under a driving force,
the obstacle can trap the pair, only one particle, or neither particle.
In Fig.~\ref{fig:8}(c) we show the collision of a pair with the obstacle
at $F_D=0.01$, where one particle becomes trapped and the other escapes.
For $F_{D} > 0.015$, the pair stays together
after encountering the obstacle, while for $F_{D}  < 0.05$, both particles become trapped.  
If the Magnus force is different in a pair of trapped particles,
two orbits form with two different average distances from the
obstacle.
Even if the two particles have Magnus forces of opposite sign, they can
still form a pinned state as shown in Fig.~\ref{fig:8}(d) for a sample with
$\alpha^1_m=2.0$, $\alpha^2_m=-2.0$, $R=1.5$, and $F^D=0.005$.

\subsection{Interaction with Multiple Obstacles and Ratchet Effects}

\begin{figure}
\includegraphics[width=3.5in]{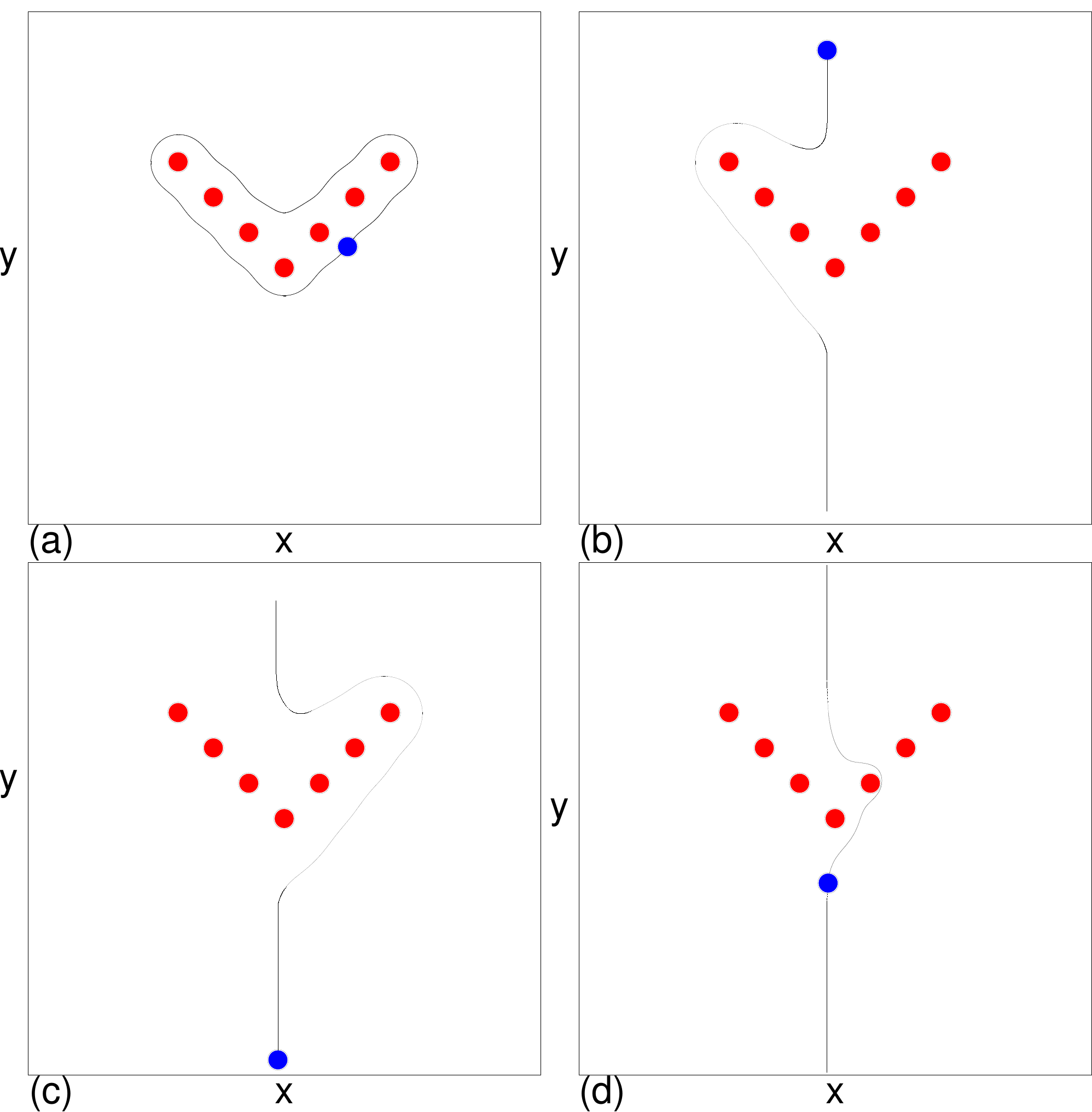}
\caption{The particle position (blue dot) and trajectory (line) along with the obstacle
  positions (red dots) for a particle with $\alpha_{m} = 1.0$
  under a drive interacting with an array of obstacles placed in a funnel configuration.
(a) At $F_{D} = 0.0$, the particle is bound to the funnel array 
and follows an orbit that encircles the obstacles in a counterclockwise direction.
(b) For a {\it negative} $x$ direction drive of $F_{D} = 0.01$,  the particle moves in the
positive $y$-direction and deviates around the obstacles.  
(c) For a drive of $F_{D} = 0.01$ applied in the positive $x$ direction,
the particle moves in the negative $y$-direction but does not
become trapped by the funnel tip. 
(d) The same as panel (c) at $F_{D} =0.250$,
where the particle passes through the funnel.  
}
\label{fig:9}
\end{figure}

When multiple obstacles are present,
a single particle can move around or encircle a cluster of obstacles to create
an edge current effect.
In an overdamped
system, when particles interact with
an asymmetric array of defects, it is possible to
create a diode effect in which the depinning
threshold is higher in one direction than the other. 
In Fig.~\ref{fig:9}(a) we show seven obstacles that have been arranged into
a funnel shape.
When a mobile particle is initially placed near one of the obstacles,
it can encircle a single obstacle or it can encircle all of
the obstacles, as shown in
Fig.~\ref{fig:9}(a) for an $\alpha_{m} = 1.0$ particle 
placed at a distance of $R = 1.5$ from the funnel, where $F_{D} = 0.0$. 
This ability to encircle multiple obstacles indicates that the
Magnus dominated particle exhibits an edge current behavior of the type
observed in chiral 
active matter systems \cite{vanZuiden16,Reichhardt19}.
Under application of a drive in the {\it negative} $x$-direction with $F^D=0.01$,
Fig.~\ref{fig:9}(a)
indicates that the particle moves in the positive $y$ direction
and curves around the array of obstacles.
The same drive of $F^D=0.01$ applied in the positive $x$ direction
causes the particle to move in the negative $y$-direction,
and as shown in Fig.~\ref{fig:9}(c),
the particle skirts around the funnel tip without getting trapped.
Under varied parameters,
we have not
found a case in which the funnel tip is able to trap the particle for driving in
any direction.
At higher $F_{D}$, the particle breaks through the funnel
array rather than moving around it, as illustrated in
Fig.~\ref{fig:9}(d) for the system
from Fig.~\ref{fig:9}(c) at $F_{D} = 0.25$.

\begin{figure}
\includegraphics[width=3.5in]{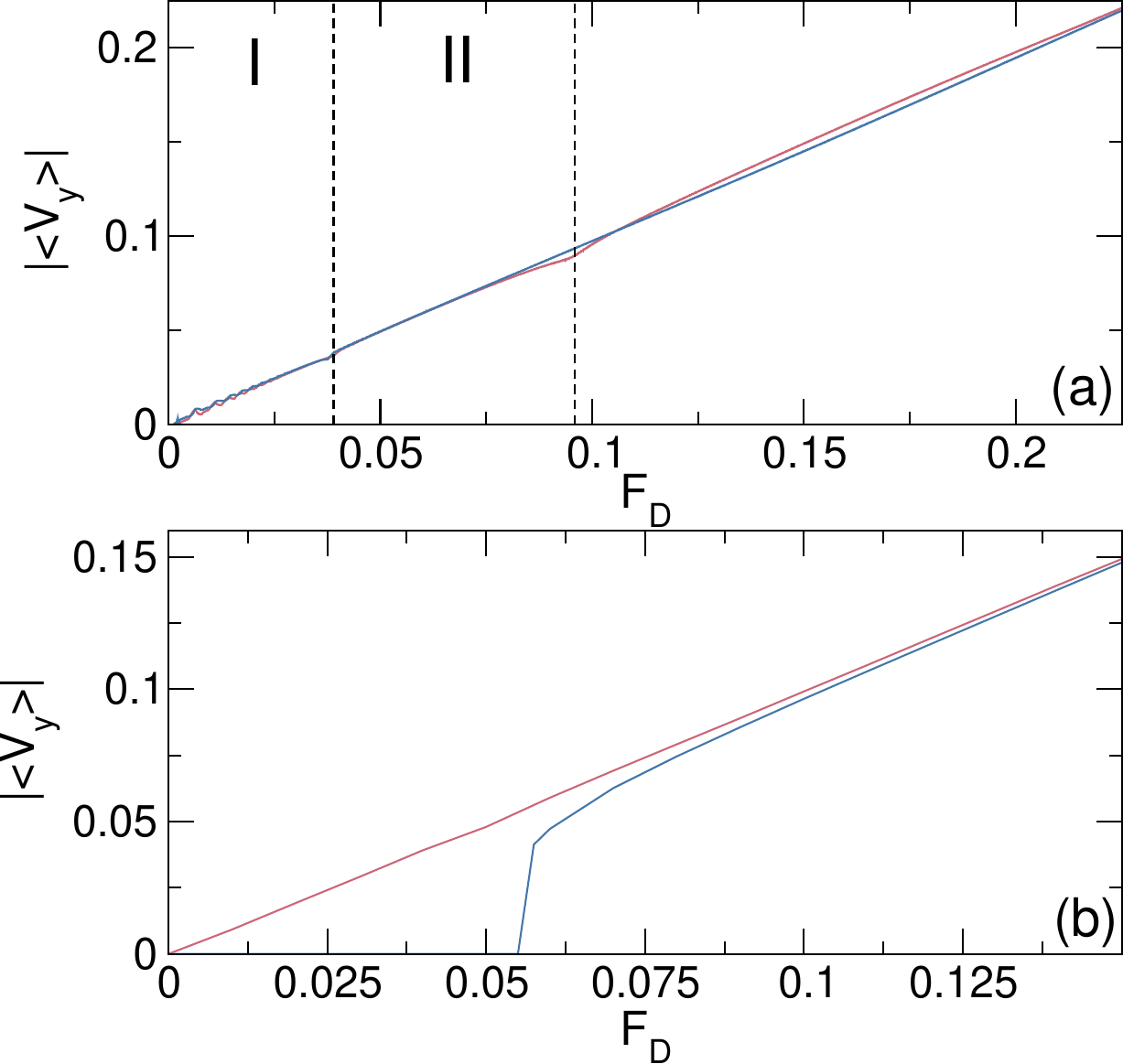}
\caption{(a) The absolute velocity $|\langle V_{y}\rangle|$ vs $F_{D}$
  for the system in Fig.~\ref{fig:9}(b,c) for motion in the
  negative $y$-direction (blue) and positive $y$-direction (pink).
  For either direction of drive,
  in Region I, the particle moves around the obstacles,
  and in Region II, the particle breaks through the funnel between the outer two
  obstacles.  The dashed line at $F_D=0.95$ indicates a transition for the
  positive $y$ direction motion to the flow illustrated in Fig.~\ref{fig:9}(d).
  Changes in the breakthrough
  location are associated with small cusps in the velocity-force curve, and additional
  breakthrough cusps occur at higher drives (not shown).
  (b) $|\langle V_y\rangle|$ vs $F_D$ for the same system in the overdamped
  limit of $\alpha_{m} = 0.0$ and $\alpha_{d} = 1.0$.
  There is a finite depinning threshold for motion in the negative
  $y$-direction (blue) but not for motion in the positive $y$ direction (pink), creating
a diode effect.   
}
\label{fig:10}
\end{figure}

In Fig.~\ref{fig:10}(a) we plot the absolute
$y$-direction velocity $|\langle V_{y}\rangle|$ versus
$F_{D}$ for the system in Fig.~\ref{fig:9} for driving in both the positive
and negative $x$ directions.
There is no pinned regime and the velocities are almost identical for both directions
of driving.
The label I indicates the regime in which the particle moves around the outer edge
of the obstacles as shown in 
Fig.~\ref{fig:9}(b,c), while the label II denotes the regime in which the particle passes
between the outer two obstacles.
For motion in the positive $y$ direction, the next breakthrough point
occurs at $F_{D} = 0.1$, which appears as a cusp in the velocity, and is associated
with a transition to the motion illustrated in Fig.~\ref{fig:9}(d).
This breakthrough transition occurs at a drive higher than the range shown
for motion in the negative $y$-direction.

\begin{figure}
\includegraphics[width=3.5in]{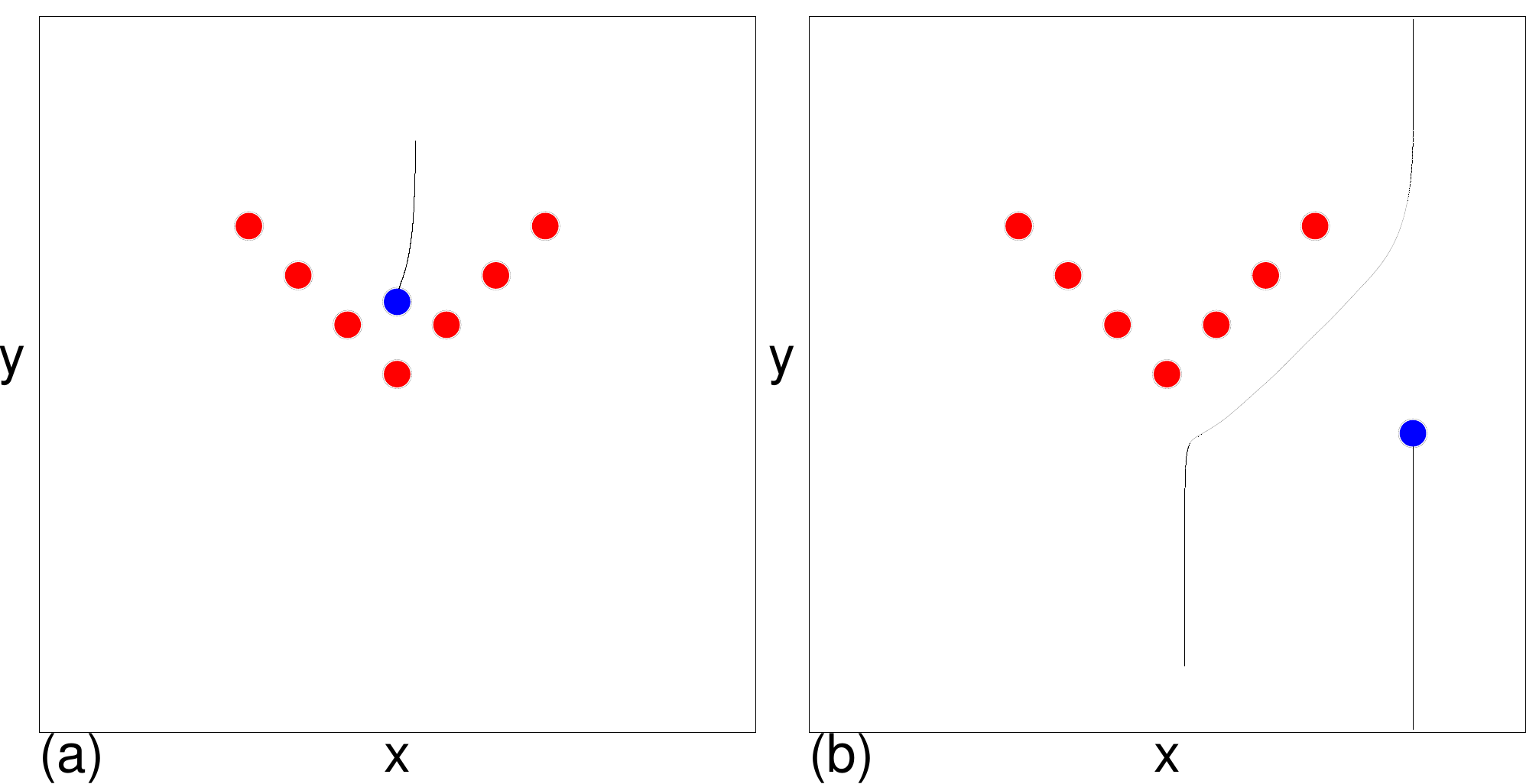}
\caption{ The particle position (blue dot) and trajectory (line) along with the obstacle
  positions (red dots) for a particle with $\alpha_{m} = 0.0$ 
  and $\alpha_{d} = 1.0$  at $F_D=0.04$ in the overdamped limit of the system in
  Fig.~\ref{fig:10}.
(a) Motion in the negative $y$-direction where the particle becomes trapped. 
  (b) Motion in the positive $y$-direction where the particle
  moves around the obstacle array and 
does not become trapped.  
}
\label{fig:11}
\end{figure}

If finite damping is present,
we can observe a diode effect which is the most pronounced in the
fully overdamped limit. 
In Fig.~\ref{fig:10}(b) we plot $|\langle V_y\rangle|$ versus
$F_{D}$ for the system from Fig.~\ref{fig:10}(a)
but with $\alpha_{m} = 0.0$ and $\alpha_{d} = 1.0$ under both positive and negative
$y$ direction driving.
Since the Magnus force is zero, the particle motion is aligned with the
driving force direction.
There is a finite depinning threshold for motion in the
negative $y$-direction, but no threshold
for driving in the positive $y$-direction. 
In Fig.~\ref{fig:11}(a) we plot the particle trajectory in the overdamped limit
of the system in Fig.~\ref{fig:10} for a drive of $F^D=0.04$
in the negative $y$-direction,
where
the particle becomes trapped by the funnel tip,
while in Fig.~\ref{fig:11}(b) the same system under driving in the positive
$y$ direction has continuous flow of
the particle around the obstacles. 

The appearance of a diode effect in the overdamped system with a
funnel array geometry also implies that if an ac drive  is applied,
a ratchet effect will appear in which the particle translates along
the easy flow direction of the funnel during one portion of 
the ac cycle.
This type of ratchet is known as a rocking ratchet \cite{Reimann02} and it
has been observed
in overdamped superconducting
vortices interacting with asymmetric pinning
\cite{Reichhardt05,deSouzaSilva06a,Lu07,Lin11,PerezdeLara11} and
in skyrmion systems where there is a combination of damping and 
a Magnus effect \cite{Reichhardt15a2,Ma17}.
In the skyrmion system there are even cases where a ratchet effect only occurs when
the Magnus force is present
\cite{Reichhardt15a2}.
The results in Figs.~\ref{fig:9} and \ref{fig:10} suggest
that if there is only a Magnus force without damping, the ratchet
effect is absent, indicating that some damping is necessary for ratcheting to occur;
however, we next show that it is still possible to achieve
a ratchet effect in the Magnus-dominated regime
if the symmetry is broken by a combination of ac
driving and the chirality of the Magnus force.

\begin{figure}
\includegraphics[width=3.5in]{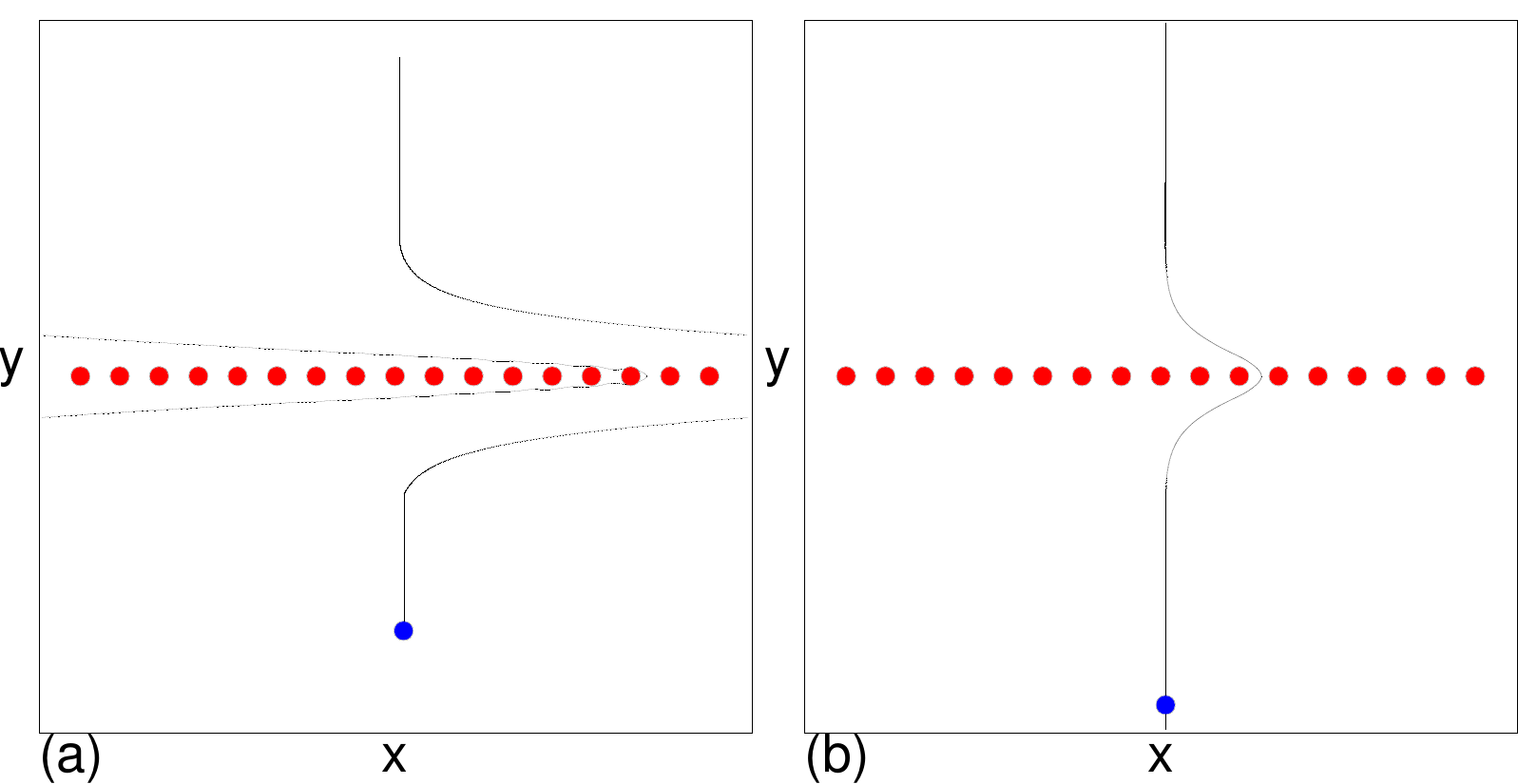}
\caption{ The particle position (blue dot) and trajectory (line) along with the
  obstacle positions (red dots) for a particle with $\alpha_{m} = 2.0$ moving toward
  a line of repulsive obstacles.
  (a) At $F_{D} = 0.007$, the particle trajectory deviates into the positive $x$
  direction as it approaches the line of obstacles.
  (b) At $F_{D} = 0.07$, the $x$-direction deviation is smaller.
}
\label{fig:12}
\end{figure}

In Fig.~\ref{fig:12}(a) we plot the trajectory of
a particle moving in the negative $y$-direction interacting with a line of obstacles
with a period of $a = 1.0$
for a system with $F_{D} = 0.007$ and $\alpha_{m} = 2.0$.
In the absence of 
obstacles, the particle
moves in a straight line at a constant velocity;
however, as the particle approaches the line of obstacles,
it begins to bend away from the line due to the
repulsive force from the particles
in the positive $y$ direction.
The Magnus force changes this repulsive force into
a positive $x$ direction velocity component of the moving particle.
The particle accelerates as it comes closer to the obstacles,
and eventually it passes through
the barrier.
As $F_{D}$ increases, the particle experiences a smaller $x$ direction
deviation of its motion when it approaches the obstacle line,
as shown in Fig.~\ref{fig:12}(b) for $F_{D} =0.07$,
while for even higher values of $F_{D}$, the deviation in the $x$-direction 
nearly disappears.
If a particle is placed near the line of obstacles in the absence of
a driving force, the
particle moves at a constant velocity parallel to the line of obstacles
due to the Magnus force.

\begin{figure}
\includegraphics[width=3.5in]{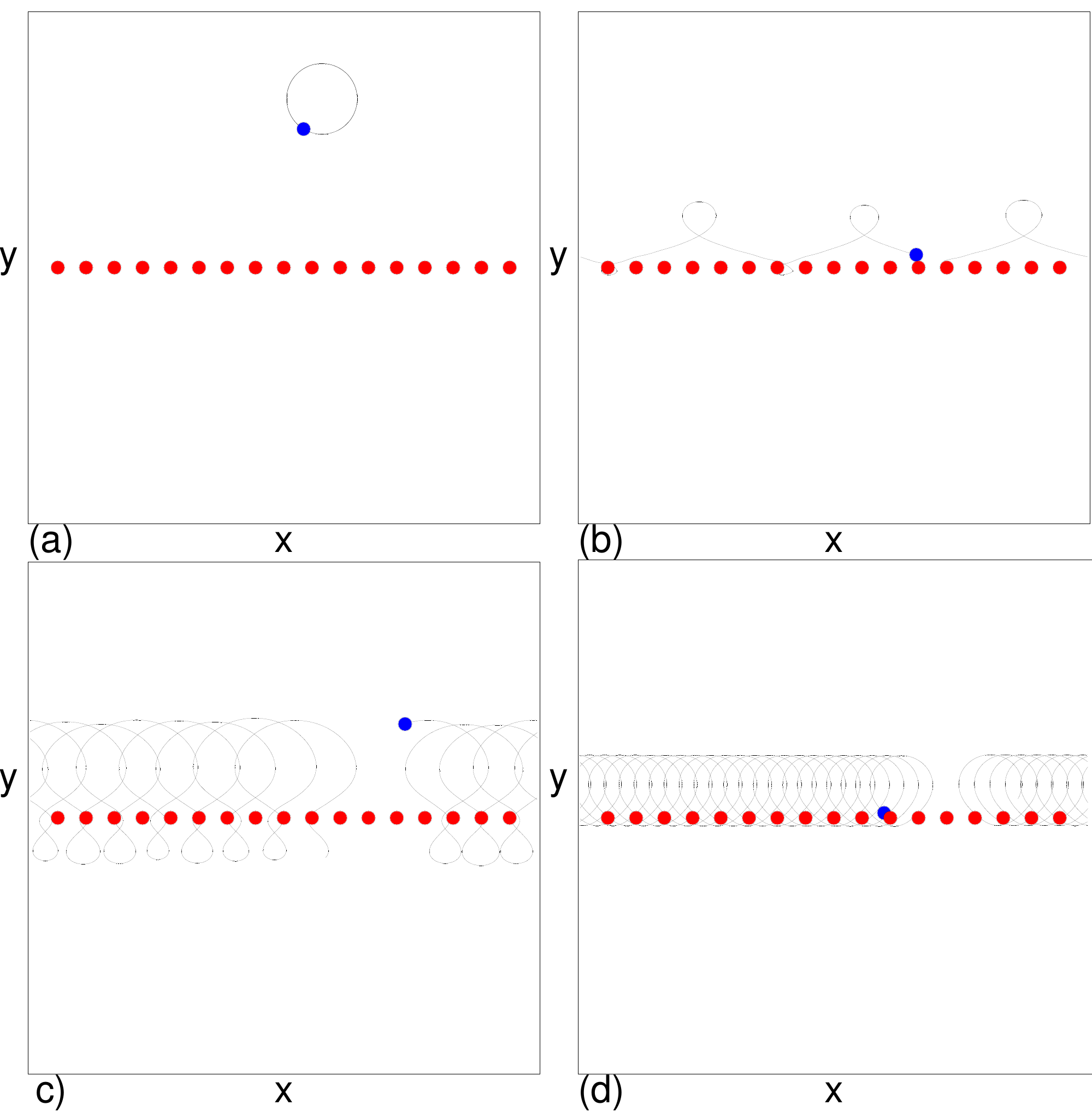}
\caption{The particle position (blue dot) and trajectory (line) along with the
  obstacle positions (red dots) for a particle interacting with a line of obstacles
  while subjected to an ac drive in the $x$ and $y$ directions.
  (a) A particle with $A = B = 0.05$, $\omega = 0.00005$,
  and $\alpha_{m} = 2.0$ placed at $R = 12a$, where there is no directed
  motion.
  (b) The same as panel (a) but with the particle placed at
  $R = 2a$, where now directed motion occurs in the positive $x$-direction.
  (c) The same as panel (b) but with $A = B =  0.1$, where the
  directed motion is
  in the negative $x$-direction.
  (d) The same as panel (b) but with $\alpha_{m} =10$, where the
  directed motion is
  in the positive $x$-direction.  
}
\label{fig:13}
\end{figure}

If we place the particle near the line of obstacles and subject it to
an ac driving force given by
$F^{AC} = A\cos(\omega t){\bf \hat{x}} + B\sin(\omega t){\bf \hat{y}}$,
we observe not only directed motion but
a reversal in the direction of motion as a function 
of the ac drive amplitude, Magnus force, and dissipation.
This occurs due to the fact that the ac drive
induces a rotation of the particle
that interacts like a gear mechanism
with the periodicity of the line of obstacles.      
In Fig.~\ref{fig:13}(a) we plot the trajectory of a particle with
$\alpha_{m} = 2.0$, $A = B = 0.05$, and $\omega =  0.00005$
which is placed at distance of $R =12a$
from the line of obstacles.  This is sufficiently far away that 
there is no interaction between the particle and the obstacles,
and the particle executes a circular
counterclockwise orbit with  no directed motion.  
In Fig.~\ref{fig:13}(b), we keep everything the same
but place the particle a distance $R = 2.0a$ from the line of obstacles.
The particle now translates in the positive $x$-direction and passes an
integer number of obstacles during each ac drive cycle.
In Fig.~\ref{fig:13}(c), the same system with $A = B = 0.1$ has a larger particle
orbit and the particle
translates in the negative $x$ direction, indicating a reversal of the current.
The effectiveness of the reversed ratchet effect is much lower,
with the particle translating at 1/4 the speed of its motion in the positive $x$ direction
in Fig.~\ref{fig:13}(b).
In Fig.~\ref{fig:13}(d), we show the system from Fig.~\ref{fig:13}(b) with
a much larger value of $\alpha_{m} = 10$.
The particle translates in the positive
$x$-direction but at a much smaller velocity.

\begin{figure}
\includegraphics[width=3.5in]{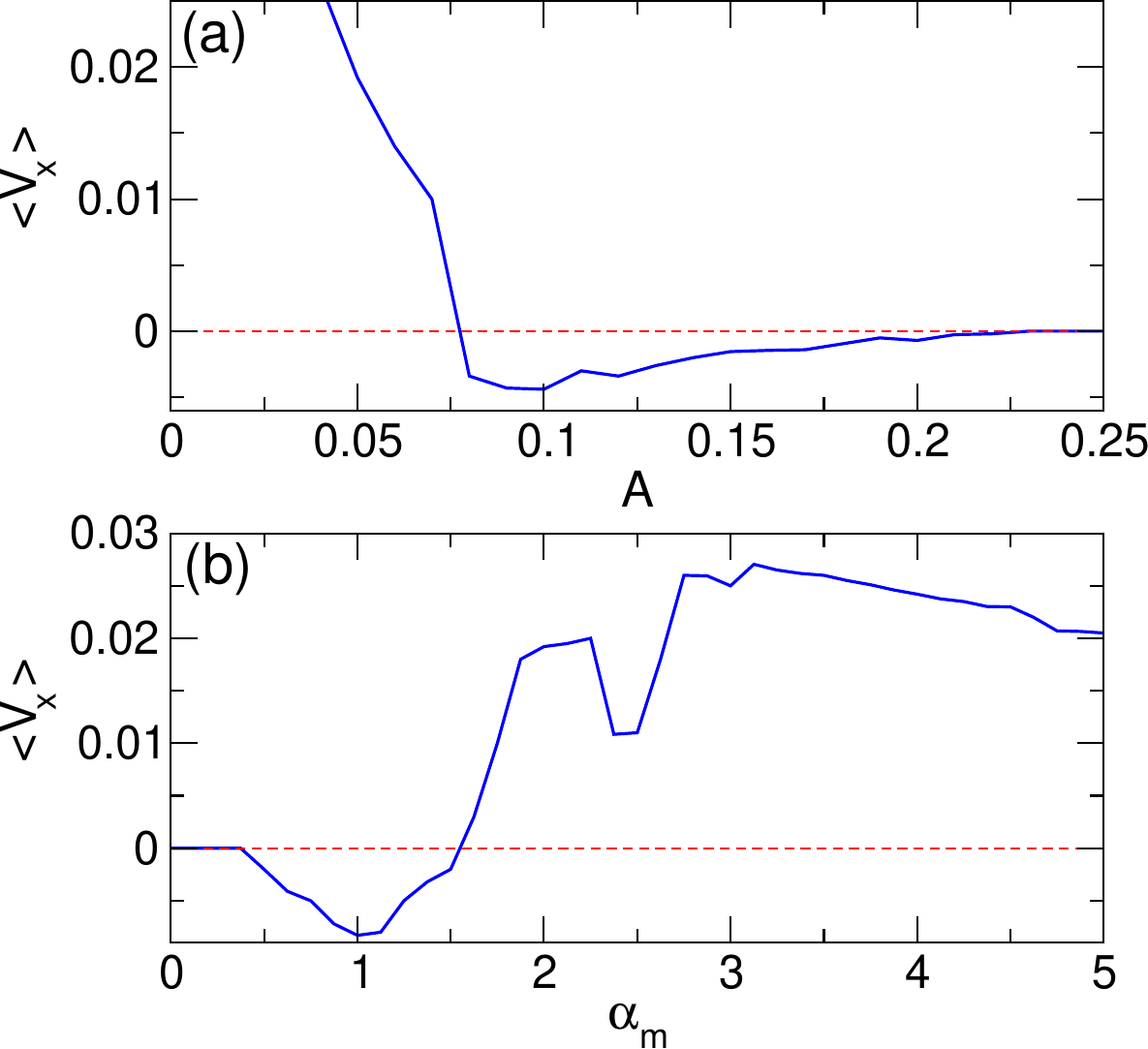}
\caption{ 
  (a) $\langle V_{x}\rangle$ vs $A$ for the system in Fig.~\ref{fig:13}(b,c)
  showing a current reversal. 
  (b) $\langle V_{x}\rangle$ vs $\alpha_{m}$ for the system in
  Fig.~\ref{fig:13}(b) with $A = 0.05$, showing 
a current reversal.  
}
\label{fig:14}
\end{figure}

\begin{figure}
\includegraphics[width=3.5in]{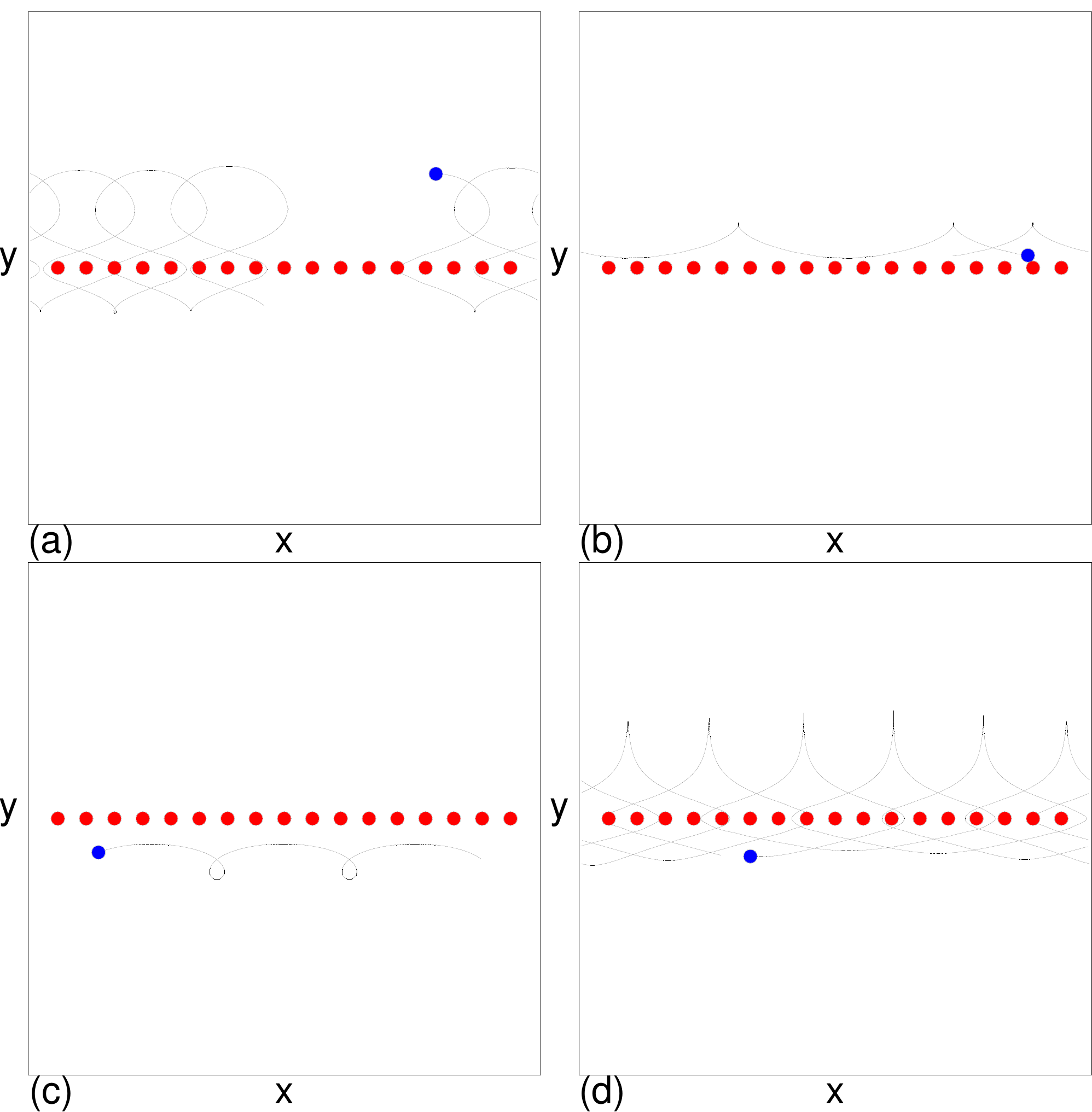}
\caption{
    The particle position (blue dot) and trajectory (line) along with the
    obstacle positions (red dots) for a particle interacting with a line of obstacles
    under an ac drive with $A=B$ and $\omega=0.00005$.
    (a) The system in Fig.~\ref{fig:14}(b) with $A=B=0.05$ at $\alpha_m=1.0$
    showing translation in the negative $x$ direction.
    (b) A translating orbit with $A=B=0.025$
    where the particle does not encircle any obstacles. 
    (c) The same as in (b) but with the particle initially placed below the line
    of obstacles, which produces translation in the negative $x$ direction.
    (d) The system from Fig.~\ref{fig:13}(b)
    with only one direction of ac drive, achieved by setting $A = 0.05$ and $B = 0.0$.
    The ratchet effect operates at only half the velocity found for simultaneous
    $x$ and $y$ driving with $A=B=0.05$.}
\label{fig:15}
\end{figure}

In Fig.~\ref{fig:14}(a) we plot $\langle V_{x}\rangle$ versus $A$ for the system
in Fig.~\ref{fig:13}(b,c) with $B=A$.  There is a reversal in the 
current from positive to negative at  $A = 0.7$,
while at higher $A$, $\langle V_{x}\rangle$ goes to zero. 
As $A$ approaches zero, the particle moves in a straight line
along the $x$ direction at fixed $\langle V_{x}\rangle = 0.056$
due to the Magnus force
created by the repulsion from the line of obstacles.  
Figure~\ref{fig:14}(b) shows $\langle V_{x}\rangle$ versus $\alpha_{m}$
for the system in Fig.~\ref{fig:13}(b) at fixed $A = 0.05$.
For $\alpha_{m} < 1.5$, the particle moves in the negative $x$ direction,
while the motion is in the positive $x$ direction when $\alpha_m\geq 1.5$.
The efficiency of the ratchet as measured by the magnitude
of $\langle V_x\rangle$ reaches a maximum near $\alpha_{m} = 3.0$
and then gradually decreases with increasing $\alpha_m$.
The step near $\alpha_{m} = 2.5$ is produced by a change in the
nature of the translating orbit.
In Fig.~\ref{fig:15}(a) we plot the trajectory of a particle moving in the
negative $x$ direction for the system in Fig.~\ref{fig:14}(b)
at $\alpha_{m} = 1.0$.
For smaller $\alpha_{m}$, the orbit increases in extent and the particle
encircles up to three obstacles
per ac drive cycle.
The magnitude and direction of the rectified current
depends on the starting position of the particle relative to the
line of obstacles,
and there can also be translating orbits that do not encircle any obstacles in which
the particle skips along the edge of the line of obstacles, as shown in
Fig.~\ref{fig:15}(b,c) for a particle with $A=B=0.025$ initially placed either above
or below the line of obstacles, respectively.
The ratchet can also occur as function of only a single ac drive.
When the ac driving force is
applied only along the $x$-direction,
we find a series of ratchet effects as illustrated in
Fig.~\ref{fig:15}(d) for the same system as in Fig.~\ref{fig:13}(b)
but with $A = 0.05$ and $B = 0.0$.
Here the
particle is ratcheting in the positive direction with $\langle V_{x}\rangle = 0.009$,
which is about half the velocity found for a ratchet effect with
simultaneous $x$ and $y$ ac driving, $A=B=0.05$.

\begin{figure}
\includegraphics[width=3.5in]{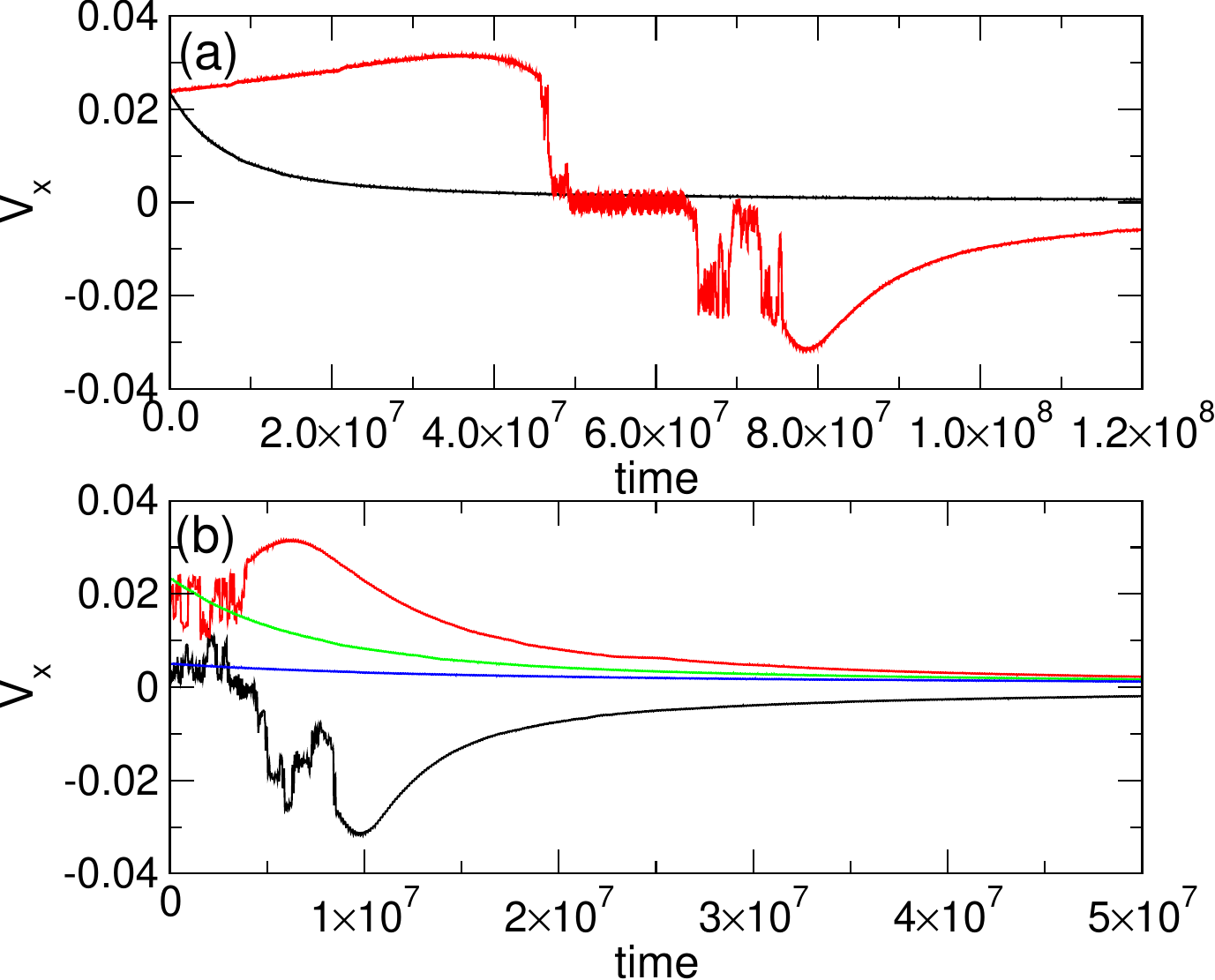}
\caption{(a) $V_{x}$ vs time in simulation time steps
  for a particle with $\alpha_m=2.0$ interacting with a line of obstacles at an
  initial distance of $R=4a$ with
  $\alpha_{d} = 0.005$ (black) and $\alpha_d=0.001$ (red).
  (b) $V_{x}$ vs time for the same system with
  $\alpha_{m} = 2.0$ and $\alpha_{d} = 0.005$ for particles initialized at
$R=a$ (black), $2a$ (red), $4a$ (green) and $6a$ (blue).  
}
\label{fig:16}
\end{figure}

The ratchet effect is strongly affected by the damping.
A finite damping term
causes a particle placed near a line of obstacles
to move away from the obstacles gradually;
however, the ac driving 
can maintain the ratcheting motion.
In Fig.~\ref{fig:16}(a) we plot $V_{x}$ versus time in simulation time steps
for a particle with $\alpha_{m} = 2.0$ and $R=4a$
at two different values of the damping, $\alpha_{d} = 0.005$ and
$\alpha_d=0.001$.
For the larger damping, the particle gradually moves in the positive $y$-direction 
away from the line of defects since the damping term aligns the
particle velocity with the direction of the repulsive force from 
the defect line.
In this case, as the particle moves further away
from the obstacles, the ratcheting effect is reduced. 
For the smaller damping, the particle oscillates across the line of
obstacles until 
it ends up below the line of obstacles
and then gradually gets pushed further away from the obstacles
in the negative $y$-direction, causing a reduction in the
ratchet effect.
In this case, there is also a window of time during which
the particle becomes localized on an
obstacle, giving a ratchet velocity of zero,
while when the particle begins to spend most of its time
below the line of obstacles, it begins to
ratchet in the negative $x$-direction.
There are also several points at which discrete jumps occur in
the velocity due to the jumping of the particle 
between different orbits that are
commensurate with the periodicity of the obstacle line.  

In Fig.~\ref{fig:16}(b), we plot $V_{x}$ versus time
in simulation time steps for the system in
Fig.~\ref{fig:16}(a) with $\alpha_{m} = 2.0$ and $\alpha_{d} = 0.005$
for a particle placed above the line of obstacles at a distance of
$R=1a$, $2a$, $4a$, and $6a$.
In this case, a particle initially placed at $R=a$ 
ends up below the line of obstacles
and is gradually pushed further in the negative $y$
direction while $V_{x}$ approaches zero. 
For $R = 2a$, the particle gradually moves away in the positive $y$-direction 
but the system passes through a series of different types of
orbits that ratchet in the positive $x$ direction,
as indicated by the oscillations in $V_x$,
and there is even a peak in the velocity before it dies away to zero.
For $R = 4a$, the 
particle enters a single orbit
and gradually moves away from the line of obstacles.
If we place the particle even further away, we observe the same behavior
as for the $R = 4a$ sample  but
with even lower values of $V_x$, as shown for $R=6a$.

We note that ratchet effects with biharmonic drives have been studied
for skyrmions, where a Magnus effect can come into play; however, in these
studies there was still a damping term, and the internal modes of the skyrmion
were also important \cite{Chen19,Chen20}.
The ratchet 
effect we observe here is more
closely related to ratchet effects found in colloids undergoing circular orbits
while interacting
with a  magnetic bubble lattice, where
the asymmetry necessary to produce the ratchet arises from the ac drive and
the transport occurs due to a commensuration effect with the underlying substrate
\cite{Chen19,Tierno07,Loehr18} 

\subsection{Dynamics of Clusters}

\begin{figure}
\includegraphics[width=3.5in]{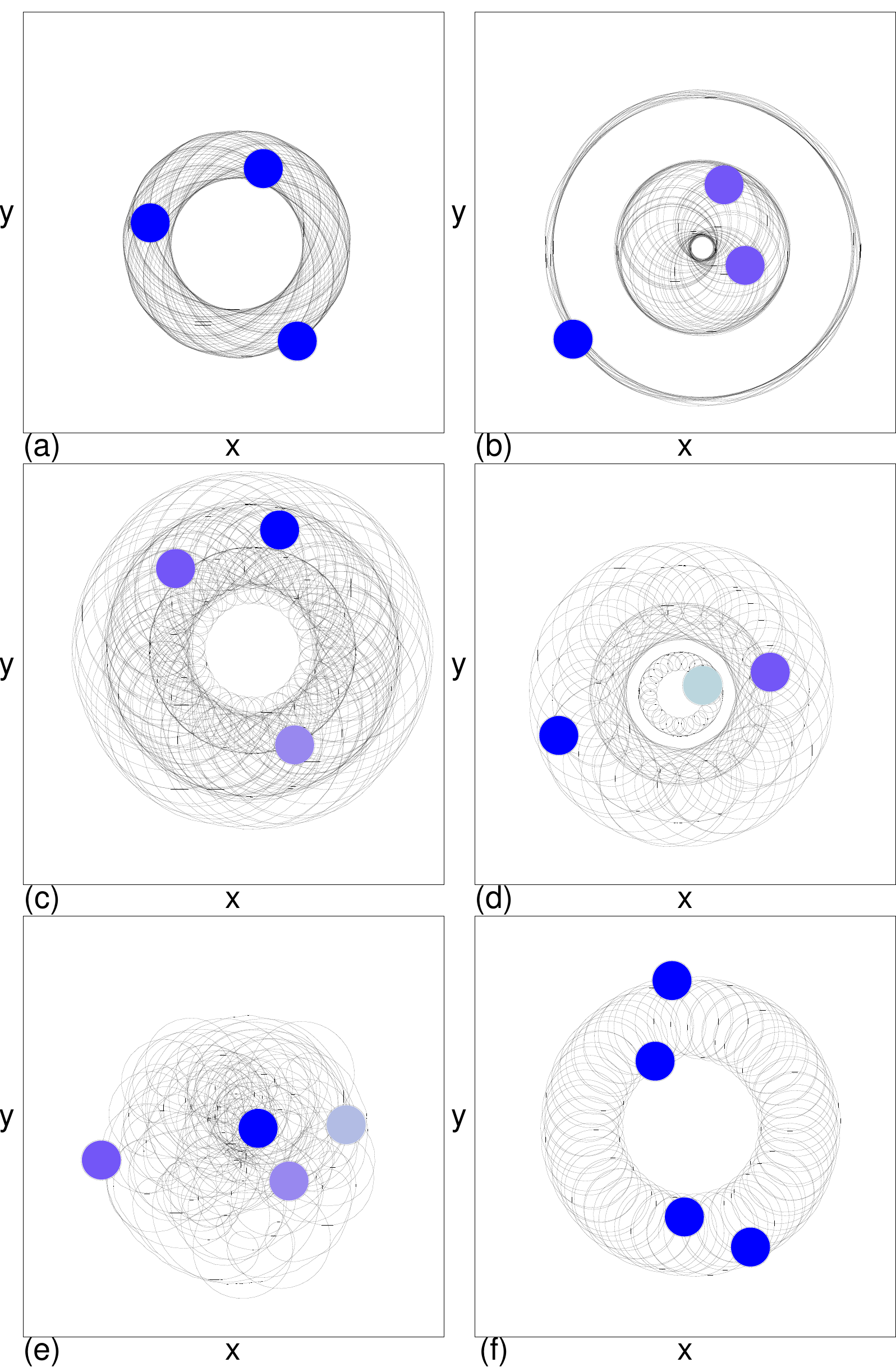}
  \caption{ The particle positions (dots) and trajectories (lines) for
    multiparticle systems with no drive.
    (a) $N = 3$, $\alpha^1_m=\alpha^2_m=\alpha^3_m = 1.0$ (blue).
    (b) $N = 3$, $\alpha^1_m=\alpha^3_m=2.0$ (blue), and
    $\alpha^2_m=1.0$ (purple).
    (c) $N = 3$, $\alpha^1_m=1.0$ (blue), $\alpha^2_m=3.0$ (light purple),
    and $\alpha^3_m=2.0$ (dark purple).
    (d) $N = 3$,  $\alpha^1_m=1.0$ (blue), $\alpha^2_m=7.0$ (light blue),
    and $\alpha^3_m=2.0$ (purple).
    (e) $N = 4$, $\alpha^1_m=1.0$ (dark blue), $\alpha^2_m=2.0$ (dark purple),
    $\alpha^3_m=3.0$ (medium purple), and
    $\alpha^4_m=4.0$ (light blue).
    (f) $N = 4$, $\alpha^1_m=\alpha^2_m=\alpha^3_m=\alpha^4_m = 2.0$ (blue). }
\label{fig:17}
\end{figure}

We next consider the case of three or more particles.
In Fig.~\ref{fig:17} we show some representative examples of possible
multiparticle orbits.
For $N=3$ particles with $\alpha^1_m=\alpha^2_m=\alpha^3_m=1.0$ that are
initially placed in a row along the $x$ direction spaced $2a$ apart,
Fig.~\ref{fig:17}(a) shows
the formation of a spiraling pattern,
which rotates due to precession of the orbits.
The particular type of orbit that appears for $N = 3$ equivalent particles
depends on the initial particle placement,
but in general we find non-chaotic stable orbits.
In Fig.~\ref{fig:17}(b) we plot the trajectories
for $N = 3$ with $\alpha^1_m=\alpha^3_m=2.0$ and $\alpha^2_m=1.0$, where
the two $\alpha_m=2$ particles form a pair 
that orbits in the center of the cluster
while the $\alpha_m=1.0$ particle follows an orbit with a larger radius. 
An $N=3$ sample in which all of the particles are different, with
$\alpha^1_m=1.0$, $\alpha^2_m=3.0$, and $\alpha^3_{m}=2.0$, appears
in Fig.~\ref{fig:17}(c).
Here a layering effect occurs in which particles with larger Magnus force
spend more time closer to the center of the cluster.
Figure~\ref{fig:17}(d) shows the same system with
$\alpha^1_m=1.0$, $\alpha^2_m=7.0$, and $\alpha^3_m=2.0$,
where three clear spatial layers appear and the $\alpha_m=7.0$ particle is
nearest to the center.
This system has some similarities the ordering
of small clusters of colloids in a trap;
however, in this case, the particles are continuously undergoing motion and there
is no external confining trap. 
In Fig.~\ref{fig:17}(e) we plot the trajectories for an
$N = 4$ system with varied Magnus forces of
$\alpha^1_{m}=1.0$, $\alpha^2_m=2.0$, $\alpha^3_m=3.0$, and $\alpha^4_m=4.0$,
which forms a chaotic cluster.
We note that if the variations in the Magnus forces are larger,
ordered states can appear 
with ring like structures, which we describe in the next subsection. 
In Fig.~\ref{fig:17}(f) we show an $N = 4$ sample
with $\alpha^1_{m}=\alpha^2_m=\alpha^3_m=\alpha^4_m = 2.0$.
In this case,
the particles form two rotating pairs which rotate around each other.
For $N >3$, most orbits are chaotic,
but for special initial placement conditions, 
it is possible to stabilize different types of rotating states. 
In larger clusters where the particles all have the same Magnus force,
the chaotic states
typically involve a
transient state of two or three particle subclusters that break up and reform over time. 

\subsection{Clusters with Strong Variation in Magnus Force Magnitude and Ring Formation}

\begin{figure}
\includegraphics[width=3.5in]{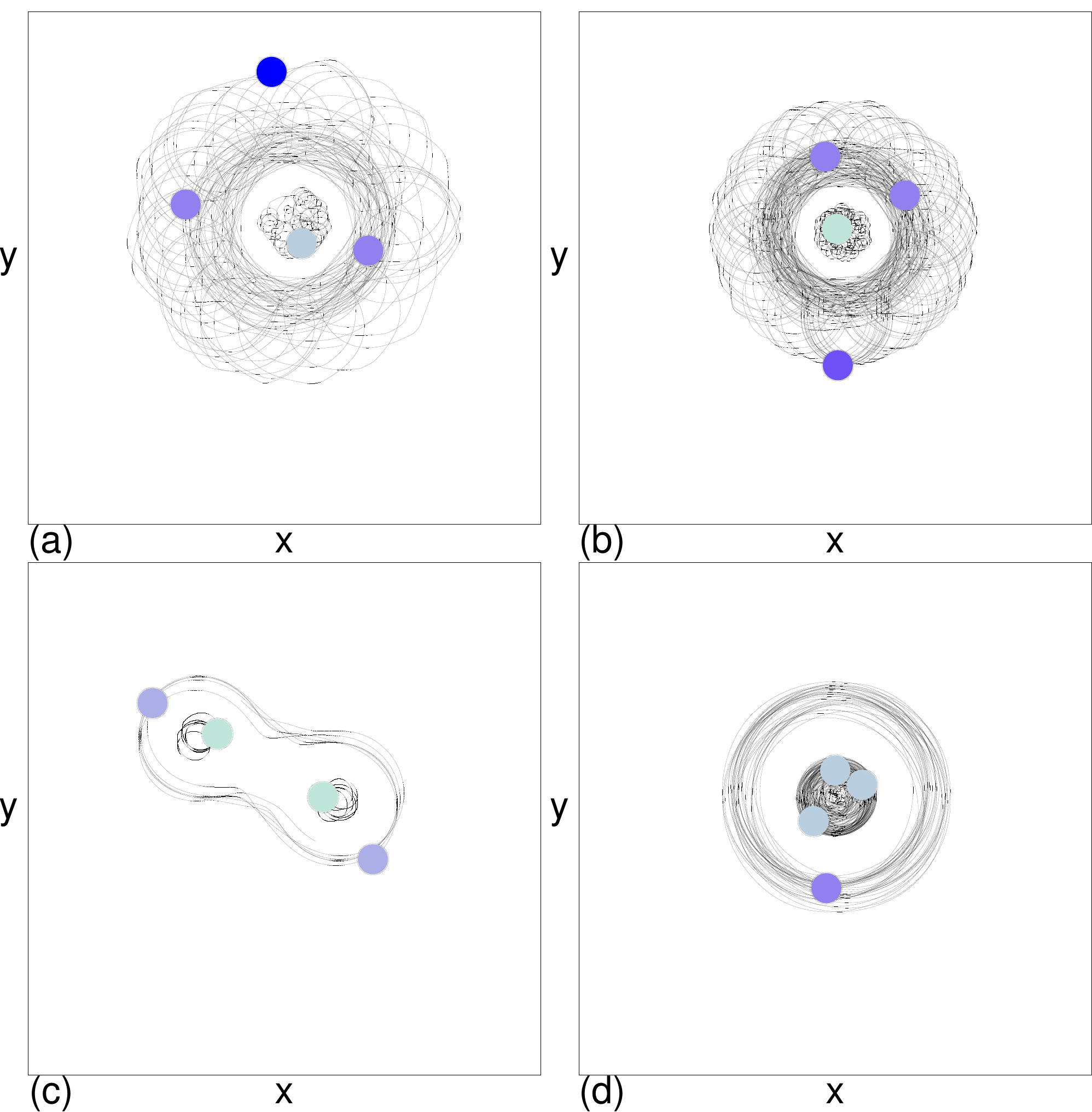}
  \caption{ The particle positions (dots) and trajectories (lines)
    showing ring like structures in multiparticle systems with strong variations in the
    Magnus forces.
    (a) $N = 4$, $\alpha^1_m=7.0$ (light blue), $\alpha^2_m=\alpha^3_m=2$ (purple), and
    $\alpha^4_m=1$ (dark blue).
    (b) $N = 4$, $\alpha^1_{m} = 10$ (light blue), $\alpha^2_m=1.5$ (dark purple),
    and $\alpha^3_m=\alpha^4_m=2$ (light purple). 
    (c) $N = 4$, $\alpha^1_m=\alpha^2_m=10$ (light blue),
    and $\alpha^3_m=\alpha^4_m=3$ (light purple),
    showing a dumbbell structure.
(d) $N = 4$, $\alpha^1_m=\alpha^2_m=\alpha^3_m=7$ (light blue), and $\alpha^4=2$ (purple).  
}
\label{fig:18}
\end{figure}

For particles with Magnus forces that are of the same sign but that have
sufficiently different magnitudes,
clusters appear
that have well defined spacings between the particle orbits,
with the particles that have the highest Magnus force
localized at the center of the cluster. 
In Fig.~\ref{fig:18}(a) we plot the trajectories in an
$N = 4$ system with $\alpha^1_m=7.0$, $\alpha^2_m=\alpha^3_m=2.0$,
and $\alpha^4_m=1.0$.  The $\alpha_m=7.0$ particle becomes localized at the
center of the cluster
and is surrounded by a ring containing the $\alpha_m=2.0$ particles,
while the $\alpha_{m} = 1.0$ particle
jumps between the $\alpha_m=2.0$ ring and a partially formed 
outer ring.
A similar structure appears in Fig.~\ref{fig:18}(b)
for an $N=4$ system with
$\alpha_{m}^1 = 10.0$, $\alpha^2_m=1.5$, and $\alpha^3_m=\alpha^4_m=2.0$.
Other cluster shapes can form for $N=4$, such as the
$\alpha^1_m=\alpha^2_m=10.0$ and $\alpha^3_m=\alpha^4_m=3.0$
system shown in Fig.~\ref{fig:18}(c)
where the two inner particles with $\alpha_m=10.0$
are orbited by the $\alpha_{m} = 3.0$
particles to form a dumbbell shape.
Strongly segregated ring structures can also occur when $N = 4$,
as illustrated in
Fig.~\ref{fig:18}(d) for a sample with
$\alpha^1_m=\alpha^2_m=\alpha^3_m=7$ and $\alpha^4_m=2$,
where the inner particles have the higher Magnus force. 
If the difference between the Magnus forces of the particles
is reduced, the ring structures are lost. 

\subsection{Clusters and Collisions for Particles with Opposite Magnus Forces}

\begin{figure}
\includegraphics[width=3.5in]{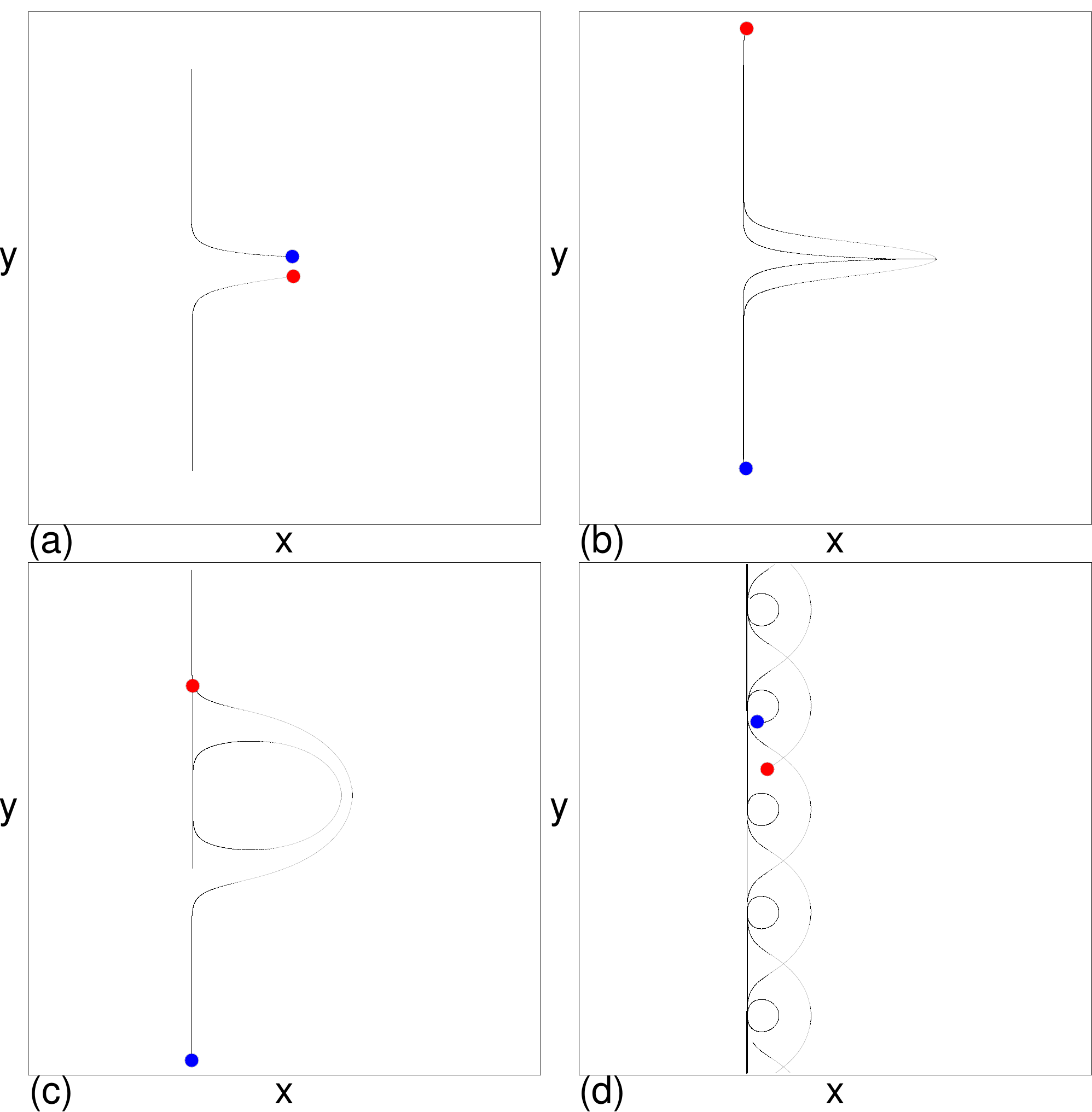}
\caption{(a,b) The particle positions (dots) and trajectories (lines) for two
  particles   initialized at opposite ends of the sample
  under a driving force of $F_{D} = 0.075$.
  (a) The first portion of the collision for $\alpha^1_m=2.0$ and $\alpha^2_m=-2.0$. 
  (b) Continuation of the motion in (a) after the particles have passed one another.
  (c) Collision for $\alpha^1_{m} = -2.0$ and $\alpha^2_{m} = 1.85$. 
(d) Collision for $\alpha^1_{m} = 2.0$ and $\alpha^2_{m} = -1.0$.   
}
\label{fig:19}
\end{figure}

As noted earlier, if two particles with equal and opposite Magnus forces
come together, they can form
a dipole that translates in a straight line.  If the magnitude of the Magnus forces
are different, an arching orbit appears instead.
In Fig.~\ref{fig:19}(a,b) we show the
trajectories of two particles with
$\alpha^1_{m} = 2.0$ and $\alpha^2_m=-2.0$ under an
external driving force of $F_{D} = 0.0075$.
The particles are initially placed at the same  $x$ position  but are widely separated in $y$.
Under the influence of the drive, the particles
initially move in opposite directions, but as they approach one another,
they form a pair that translates in the positive $x$ direction, 
as shown in Fig.~\ref{fig:19}(a).
The driving force causes the particles to move closer together and eventually
pass each other as shown in Fig.~\ref{fig:19}(b).
Figure~\ref{fig:19}(c) shows two particles with
$\alpha^1_{m} = 2.0$ and $\alpha^2_{m} = -1.5$ that form a dipole
which moves in an arch shape before the particles decouple again.
In Fig.~\ref{fig:19}(d),
a system with $\alpha^1_{m} = 2.0$ and $\alpha^2_m=-1.0$
undergoes multiple collisions 
due to the periodic boundary conditions,
and the orbit performed during each collision has a small radius due to
the large difference in the magnitude of the Magnus forces.
If the particles are separated in $y$ but also have a small offset
in $x$, they do not collide head on, which creates 
spiraling orbits similar to that
shown in Fig.~\ref{fig:19}(b) but with asymmetric loops.

\begin{figure}
\includegraphics[width=3.5in]{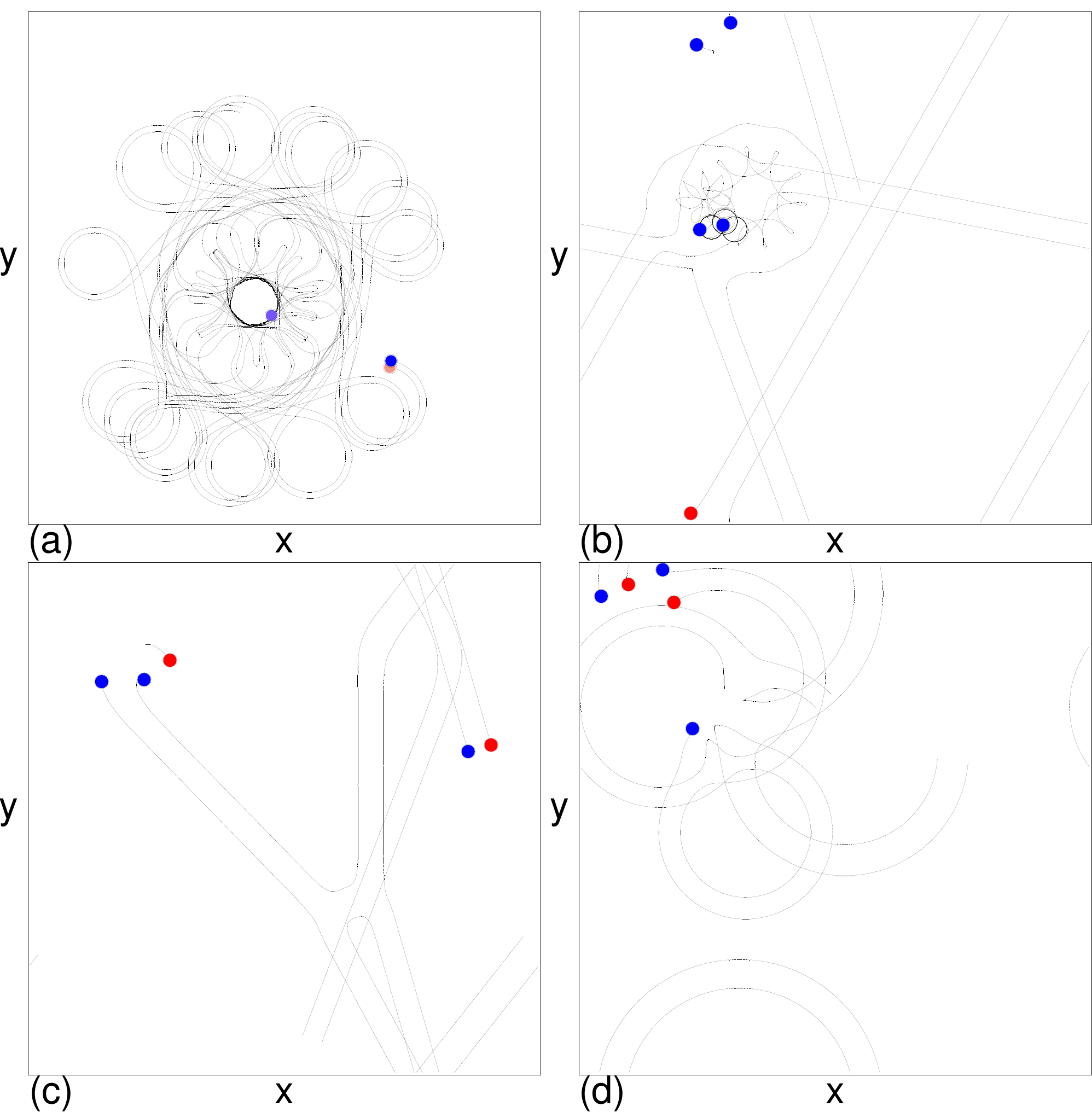}
\caption{ The particle positions (dots) and trajectories (lines) in systems
  with mixed Magnus force sign and no drive.
  (a) A closed orbit at $N = 3$, $\alpha^1_m=1.0$ (purple),
  $\alpha^2_m=-1.1$ (orange), and $\alpha^3_{m} = 0.85$ (blue).
  (b) A translating dipole at $N = 5$,
  $\alpha^1_m=\alpha^2_m=\alpha^3_m=\alpha^4_m=2.0$ (blue),
  and $\alpha^5_m=-2.0$ (red).  The two particles that are paired into the dipole
  are at the bottom and top of the image due to the periodic boundary conditions.
  (c) At $N = 5$, $\alpha^1_m=\alpha^3_m=\alpha^5_m=2.0$ (blue), and
  $\alpha^2_m=\alpha^4_m=-2.0$ (red),
  there are two intermittently  forming pairs of dipoles.
  (d) Circular dipole motion at $N = 5$,
  $\alpha^1_m=\alpha^2_m=\alpha^5_m=2.0$ (blue),
  and $\alpha^3_m=\alpha^4_m=-1.5$ (red).
}
\label{fig:20} 
\end{figure}

\begin{figure}
\includegraphics[width=3.4in]{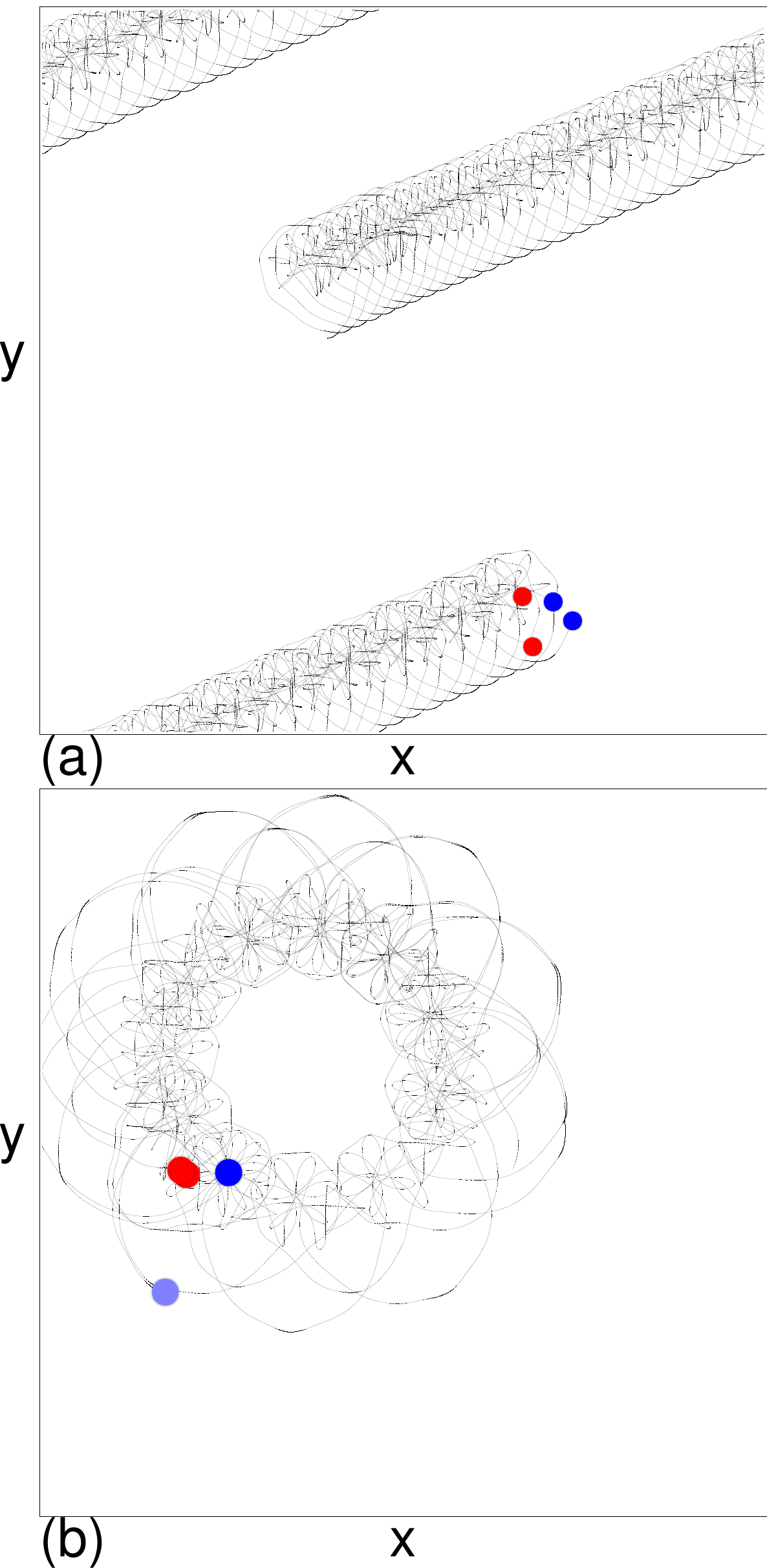}
\caption{The particle positions (dots) and trajectories (lines)
  in systems with mixed Magnus force sign and no drive.
  (a) For $N = 4$, $\alpha^1_m=\alpha^3_m=2.0$ (blue),
  and $\alpha^2_m=\alpha^4_m=-4.0$ (red),
  the particles form a translating cluster. 
  (b) For $N=4$, $\alpha^1_m=1.0$ (blue),
  $\alpha^2_m=\alpha^4_m=-2$ (red), and $\alpha^3_m=2$ (purple),
  the cluster moves in a circle.    
}
\label{fig:21}
\end{figure}

For a system of three
particles in which the sign of the Magnus term of one particle is opposite from
that of the other two particles,
we generally observe closed periodic orbits;
however, depending
on the initial placement of the particles,
it is also possible to have a pair of particles with
opposite signs of Magnus force break off and move away as a dipole. 
In a system with mixed Magnus force amplitudes
where one particle has a positive Magnus force and the other two have
negative Magnus forces,
a translating dipole can form that then rotates around the third particle.
For example, 
in Fig.~\ref{fig:20}(a), 
a system with $\alpha^1_m=1.0$, $\alpha^2_m=-1.1$, and $\alpha^3_{m} = 0.85$
has a translating dipole moving in an orbit that gradually precesses counterclockwise
while the third particle follows a tighter precessing orbit. 
For five or more particles with mixed
Magnus force signs, in general we do not observe
long-lived localized structures but instead
find that
pairs of particles with opposite sign
form a gas of translating dipoles 
that are either broken up or 
deflected when a collision with another particle or dipole occurs.
In Fig.~\ref{fig:20}(b) we show the trajectories of a system with five particles
where $\alpha^1_m=\alpha^2_m=\alpha^3_m=\alpha^4_m=2.0$ and
$\alpha^5_m= -2.0$.
One translating dipole appears, while the other particles of 
the same sign form rotating clusters.
When the
dipole encounters a rotating cluster, it typically scatters off in a new direction
after partially encircling the cluster, but there can also be an exchange of one of
the dipole particles with one of the cluster particles.
In Fig.~\ref{fig:20}(c), an $N=5$ system with
$\alpha^1_m=\alpha^3_m=\alpha^5_m=2.0$ and $\alpha^2_m=\alpha^4_m=-2.0$ has
similar dynamics, but there are now two translating dipoles which undergo
two types of collisions.
The first is the scattering of a dipole by an isolated particle, as shown in the
upper left hand portion of the figure.  The dipole can either exchange one of
its particles with the isolated particle or simply be deflected.
The second collision is a dipole-dipole scattering  in which
the dipoles can exchange particles and/or change their directions of motion.
The $N=5$ sample
with  $\alpha^1_m=\alpha^2_m=\alpha^5_m=2.0$ and
$\alpha^3_m=\alpha^4_m=-1.5$
in Fig.~\ref{fig:20}(d)
also contains two translating dipoles,
but since the Magnus forces in the dipoles are not of equal magnitude,
the dipole pairs
move in circular 
paths and can break up or be deflected when they
collide with each other or with the remaining stationary particle. 
For $N=6$ and higher,
we observe only translating and chaotic orbits. 
When $N = 4$, it is possible for the system to form a larger scale translating cluster
instead of a dipole, as shown 
in Fig.~\ref{fig:21}(a)
for $\alpha^1_m=\alpha^3_m=2$ and $\alpha^2_m=\alpha^4_m=-2$.
The cluster is composed of particles that
continuously switch between
forming pairs of the same sign that rotate and
forming pairs of the opposite sign that translate.
For this combination of Magnus forces,
we always observe translating clusters,
but the direction and velocity of the translation
depends on the initial placement of the particles.
If the Magnus forces are unequal, as in Fig.~\ref{fig:21}(b) where
$\alpha^1_m=1$, $\alpha^2_m=\alpha^4_m=-2$, and $\alpha^3=2$,
similar dynamics occur but the cluster moves in a circle.

\begin{figure}
\includegraphics[width=3.5in]{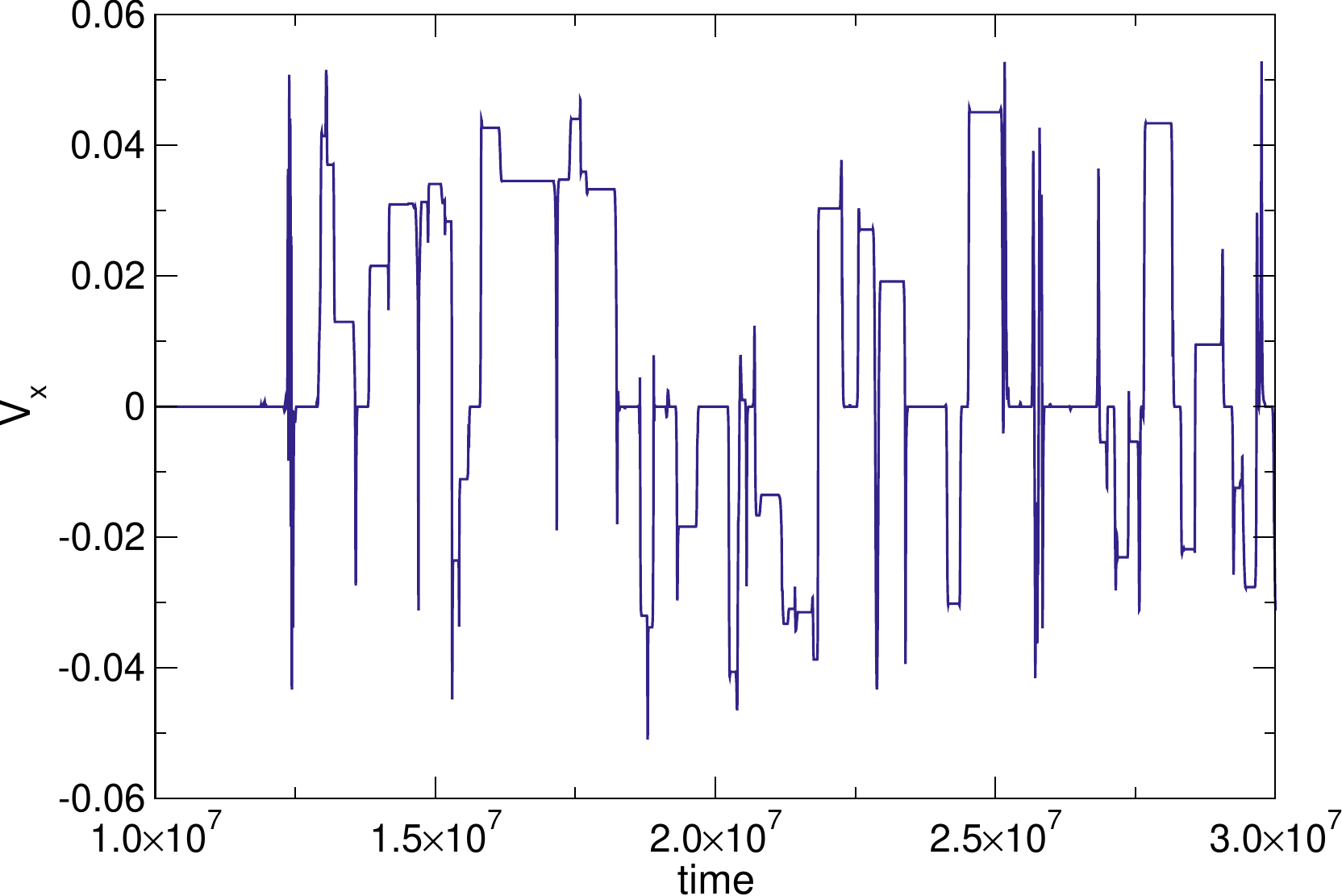}
\caption{ The velocity component $V_{x}$ vs time
  in simulation time steps for one of the particles in
  Fig.~\ref{fig:20}(c) showing jumps
  between zero velocity  when the particle is not part of a dipole
  and finite velocity when the particle is part of a translating dipole.
}
\label{fig:22}
\end{figure}

A collection of particles with opposite
Magnus force signs can be considered an example of an active matter 
system.
In active matter,
the particles are self-propelled
and can be described as undergoing driven Brownian diffusion or
run and tumble dynamics.
Typically, active
particles show short time ballistic
behavior and long time diffusive behavior due to collisions
\cite{Marchetti13,Bechinger16}.
In the case of the Magnus dominated system,
mixtures of opposite Magnus force signs 
form translating dipoles that act like active Brownian particles in the limit of zero
orientational diffusion
or like run and tumble particles with an infinite  run time.
When there are other particles in the system,
collisions can cause the dipoles to change directions or to break up before
reforming again.
To highlight this effect,
in Fig.~\ref{fig:22} we plot a time series of the $x$-direction velocity $V_x$
for a single particle from the system in
Fig.~\ref{fig:20}(c), where
regions of constant velocity are interspersed with regions of zero velocity.
The constant velocity regions correspond to periods in which the particle forms
half of a translating dipole, while the zero velocity regions are periods in which
the particle is no longer paired into a dipole and is therefore stationary.
There can also be intervals in which the particle is part of a rotating pair
composed of two particles with the same Magnus force sign.
In future studies, it would be interesting to examine
the velocity distributions in
large collections of mixtures of Magnus dominated particles
to see whether this system exhibits further similarities to active matter.

\section{Discussion}
A number of the results we observe are similar to behavior
found in point vortex models.
In these models, vortices in fluids 
are represented as non-dissipative point particles with
a logarithmic long range interaction and
nondissipative dynamics that are controlled by a Coriolis or Magnus term
\cite{Aref88,Boatto99,Aref07,Reinaud18}. 
A pair of point vortices
with the same vorticity rotate around one another,
while a pair
with opposite vorticity translates.
Additionally, the point vortex literature shows that
clusters of four or more particles generally form chaotic states. 
Other work has shown
that point vortex particles can effectively be trapped in orbits
around defects such as a fixed point vortex \cite{Ryzhov13,Koshel18},
or they can scatter off defects. 
In our case, the interactions are shorter range than the
point vortex interactions;
however, the smooth behavior of the Bessel function potential
causes much of the dynamics of the Magnus-dominated particles to be
fairly similar to the point vortices.
In our work we considered scattering off multiple objects,
ratchet effects, and particles with mixed Magnus force values.
In most of the point vortex literature,
the Magnus force is of the same magnitude, and in general, there is no driving force 
and dissipative effects are neglected.

Due to the dynamical nature of the states we observe,
it is possible to imagine that for periodic obstacle geometries
or arrangements of a large assembly of particles
localized around an obstacle,
some sort of dynamical but repeatable crystal could form
which would be an example of a classical time crystal
\cite{Shapere12,Heugel19,Yao20,Libal19}.
In a real system, some form
of dissipation would likely arise that would eventually destroy the crystal,
but it may be possible to create long lived transient
Magnus time crystals.   

Experimentally, our system most closely resembles
skyrmion or meron motion with no or low dissipation
where the Magnus force dominates the dynamics,
which should be achievable under 
certain conditions.
Skyrmions can also be set into motion readily under a drive,
so it should be possible to maintain the transient 
Magnus force dynamics
in a low dissipation system indefinitely by applying ac or dc driving.
It is also possible to have dispersion in the Magnus force component of a
skyrmion system
as well as skyrmions with opposite signs. 
Additional internal modes can arise in skyrmions that
are not taken into account in our model;
however, there have already
been some studies of skyrmion dynamics in the zero dissipation limit 
with both Thiele equation and continuum modeling approaches \cite{Feilhauer19}. 
 
Our results could also be relevant to
spinning charged colloids levitated acoustically or
dusty plasmas in magnetic fields, where Magnus effects arise and dissipation
effects are weak.
Many active spinner systems include strong dissipation or have
only short range contact interaction forces,
so that two particles or a particle and an obstacle would only interact when they touch.
Finally, our results for decoupling and depinning should also be
applicable to vortices and point vortex models
in the presence of some form of flow field.

\section{Summary} 
We have examined the dynamics of individual pairs and small
clusters of repulsive pairwise interacting particles
in which the dynamics is dominated by a Magnus term.
In the overdamped limit,
clusters of such particles exhibit transient motion and settle into a stationary state.
For particles without dissipation, when the Magnus terms have the same
magnitude and sign,
a pair of repulsively interacting particles
rotate around each other at fixed distance.
Similar rotating clusters appear up to sizes of $N=4$, but
for larger clusters the dynamics become chaotic. 
A pair of particles with opposite Magnus force sign forms a translating dipole.
Under an applied drive, an individual
particle moves at $90^\circ$ with respect to the drive
direction, a rotating pair with the
same Magnus force translates,
and a pair with different Magnus force magnitudes
has a decoupling driving force threshold.
A particle interacting with repulsive obstacles 
forms a bound state with a critical driving threshold for the decoupling
of the particle from the obstacle,
while if the particle dynamics include damping,
the particle gradually spirals away from the obstacle.
A single obstacle can bind multiple particles simultaneously.
When a rotating pair encounters a obstacle,
one or both particles in the pair can become trapped.
For particles interacting with clusters of obstacles,
we find that it is possible for a particle
to become bound to the cluster and form
a circulating current around the outside of the cluster.
In the overdamped limit, a particle 
interacting with obstacles arranged in a funnel shape
exhibits a diode effect,
but when there is only a Magnus force and no damping, the
diode effect disappears.
A line of obstacles causes a deviation in the direction of the trajectory
of the driven particle, which eventually passes through the obstacle line.
Under an ac drive, we show that
it is possible to observe a ratchet effect
for a particle
placed
near a line of obstacles due to a gear-like mechanism in which the
particle orbit becomes commensurate
with the periodicity of the obstacle line.
The ratchet effect shows a reversal as a function
of ac drive, Magnus force, and distance from the obstacle line.
For large clusters of particles, we 
find that if the dispersion in the Magnus force
is sufficiently large, the particles with the largest Magnus
force become localized in the center of the cluster.
In mixtures of particles with opposite signs, we find the intermittent
formation of dipoles
that can translate over some distance before 
breaking up or deflecting upon encountering other particles, and we show that
these dipoles
have certain similarities to active matter systems. 
Our results could applied to skyrmion systems
in the absence of dissipation or in the low dissipation limit,
or to chiral active matter
in which there is low damping or continuous driving.
Our results could also be useful for understanding transient dynamics in
systems with Magnus dominated dynamics and weak damping.

\acknowledgments
This work was supported by the US Department of Energy through
the Los Alamos National Laboratory.  Los Alamos National Laboratory is
operated by Triad National Security, LLC, for the National Nuclear Security
Administration of the U. S. Department of Energy (Contract No.~892333218NCA000001).

\bibliography{mybib}
\end{document}